\documentclass[12pt]{article}
\usepackage{bm}
\usepackage{amsmath}
        \numberwithin{equation}{section}
\usepackage{amsmath,amssymb}
\usepackage{bbold}
\usepackage{multirow}
\usepackage{enumerate}
\usepackage{ulem}
\usepackage{color}
\newcommand{\pa}{\partial}

\newcommand{\tx}{\tilde{x}}

\newcommand{\tm}{\tilde{m}}
\newcommand{\ty}{\tilde{y}}
\newcommand{\tp}{\tilde{p}}
\newcommand{\tu}{\tilde{u}}
\newcommand{\ts}{\tilde{s}}
\newcommand{\tv}{\tilde{v}}

\newcommand{\tc}{\tilde{c}}
\newcommand{\tC}{\tilde{C}}
\newcommand{\tG}{\tilde{G}}

\newcommand{\ep}{\epsilon}

\definecolor{refkey}{rgb}{1,0,0}
\definecolor{labelkey}{rgb}{0,0,1}
\begin{document}

\begin{titlepage}
\begin{flushright}
TIT/HEP-678 \\
January, 2020
\end{flushright}
\vspace{0.5cm}
\begin{center}
{\Large \bf
Quantum Seiberg-Witten periods for 
$\mathcal{N}=2$ $SU(N_c)$ SQCD 
around the superconformal point
}

\lineskip .75em
\vskip 2.5cm
{\large  Katsushi Ito, Saki Koizumi and Takafumi Okubo }
\vskip 2.5em
 {\normalsize\it Department of Physics,\\
Tokyo Institute of Technology\\
Tokyo, 152-8551, Japan}
\vskip 3.0em



\end{center}
\begin{abstract}
We study the quantum Seiberg-Witten periods of ${\cal N}=2$ superconformal field theories  which are obtained by taking the scaling limit of ${\cal N}=2$ $SU(N_c)$ SQCD around the superconformal fixed point.
The quantum Seiberg-Witten curves of these superconformal field theories 
are shown to be classified into  the Schr\"odinger type and the SQCD type, which depend on flavor symmetry at the fixed point.
We study the quantum periods and compute the differential operators which relate the quantum periods to the classical ones up to the fourth-order in the deformation parameter.

\end{abstract}
\end{titlepage}
\baselineskip=0.7cm
\section{Introduction}
The low-energy description of ${\cal N}=2$ supersymmetric gauge theories has singularities in the strong coupling region \cite{Seiberg:1994rs,Seiberg:1994aj}. 
In particular,  there appears an IR fixed point at a locus of the Coulomb branch, where mutually non-local BPS particles become massless \cite{Argyres:1995jj,Argyres:1995xn,Eguchi:1996vu}.
The dynamics around the fixed point is described by an interacting ${\cal N}=2$ superconformal field theory called the Argyres-Douglas (AD) theory.

One can study the BPS spectrum of the AD theory from the Seiberg-Witten (SW) curve,
which is  obtained  by  degeneration of the  SW curve of the UV gauge theory.
It is also realized from the compactification of the six-dimensional $\mathcal{N}=(2,0)$ superconformal field theory on a punctured Riemann surface \cite{Gaiotto:2009we,Gaiotto:2009hg,Xie:2012hs} and 
classified in \cite{Cecotti:2010fi,Wang:2015mra,Wang:2016yha,Argyres:2015ffa}. 

One can also study the SW theories in the Nekrasov-Shatashvili (NS) limit 
of the Omega-background \cite{Nekrasov:2009rc}. 
The Omega background deforms four dimensional spacetime by the two-dimensional torus action \cite{Moore:1997dj}. 
The NS limit is defined by the limit where one of the deformation parameters goes to zero.
In this limit, the deformed theory is reformulated by the quantum SW curve \cite{Mironov:2009uv},
which is a differential equation obtained by the quantization of the symplectic structure based on the SW differential. 
The deformation parameter plays a role of the Planck constant.

The WKB analysis of the quantum SW curve determines the deformed SW periods\cite{Basar:2015xna,Kashani-Poor:2015pca,Ashok:2016yxz}.
In the weak coupling region, the quantum corrections to the SW periods agree with those obtained from the NS limit of the Nekrasov partition function \cite{Mironov:2009uv,Mironov:2009dv,Zenkevich:2011zx,Popolitov:2013ria}.
In the strong coupling region such as the massless monopole/dyon  point, the quantum SW periods for the $SU(2)$ SQCD with $N_f \leq 4$ have been studied  in \cite{He:2010xa,Ito:2017iba,Beccaria:2016wop}.
The relation to the topological strings have been studied in \cite{Huang:2012kn,Krefl:2014nfa}.
(See also \cite{Maruyoshi:2010iu,Kanno:2013vi,Piatek:2014lma,Poghossian:2016rzb} for its relation to CFT.)

In the previous papers \cite{Ito:2018hwp,Ito:2019twh},  we have studied the quantum SW curves for the AD theories associated with  ${\cal N}=2$ $SU(2)$ SQCD, the $(A_1,A_r)$-type  and $(A_1,D_r)$-type AD theories.
We have calculated the quantum corrections to the SW periods  up to the fourth order in the deformation parameter and confirmed that they are consistent with the scaling limit of the quantum periods of the UV theories.
Recently the exact WKB periods for the quantum SW curve of $(A_1,A_r)$-type AD theory are shown to be determined by  the Thermodynamic Bethe ansatz (TBA) equations \cite{Gaiotto:2014bza, Ito:2018eon}. 
Moreover the periods satisfy the perturbative-non-perturbative (PNP) relations.
It is an interesting problem to generalize  to other quantum SW curves and to study their integrable structure. See for example \cite{Ito:2017ypt,Grassi:2019coc,Fioravanti:2019vxi,Ito:2019llq} for recent developments.

The purpose of this paper is to study the quantum SW curve and the quantum SW periods for the AD theories obtained from the scaling limit of ${\cal N}=2$ $SU(N_c)$ SQCD  around the superconformal point.
These AD theories are basic examples of conformal field theories with various order parameters and 
flavor symmetries \cite{Eguchi:1996vu,Gaiotto:2010jf}.
It will be shown that the quantum SW curve can be classified into the Schr\"ordinger type differential equation or the quantum SW curve of SQCD type, which depends on the flavor symmetry.
We will calculate 
the quantum corrections to the classical SW periods from the WKB expansions of a solution of the quantum SW curve. 
They are also obtained from the SW periods of the corresponding UV theories by taking the scaling limit.
The quantum corrections are obtained by acting the differential operators to the classical SW periods.
We will confirm that  the quantum  SW periods of the AD theories are consistent with  the scaling limit of the $SU(N_c)$ SQCD up to fourth-order in $\hbar $.

This paper is organized as follows:
in section 2, we review the SW curves and the SW differentials for the AD
 theories which are obtained by the scaling limit of ${\cal N}=2$ $SU(N_c)$ SQCD.
In section 3, we discuss the quantum SW curves of  ${\cal N}=2$ $SU(N_c)$ SQCD and the  related AD theories.
In section 4, we study the quantum SW periods for the AD theories. We calculate the second and fourth-order corrections to the classical periods.
Section 5 is devoted to conclusions and discussion.

\section{Seiberg-Witten curve of $SU(N_c)$ SQCD around the  superconformal point }
We start with introducing
the SW curve for 
${\cal N}=2$ 
$SU(N_c)$ gauge theory with $N_f$ hypermultiplets ($N_f< 2N_c$) as \cite{Hanany:1995na}:
\begin{align}
y^2=&C(x)^2-G(x),
\label{eq:swc}
\end{align}
where 
\begin{align}
C(x)=&
x^{N_c}-\sum ^{N_c}_{i=2} s_i x^{N_c-i},
 \label{eq:swcC}\\
G(x)=&\Lambda^{2N_c-N_f} \prod _{a=1}^{N_f} (x+m_a). \label{eq:swcG}
\end{align}
Here $s_i$ $(i=2,\cdots ,N_c)$ are the Coulomb moduli parameters, $m_a$ $(a=1,\cdots ,N_f)$ are the mass parameters of $N_f$ fundamental hypermultiplets, $\Lambda$ is the QCD scale parameter.
The SW differential is given by
\begin{align} \label{swdiff}
\lambda _{\text{SW}} =x d\log \frac{C(x)-y}{C(x)+y}.
\end{align}
The derivative of $\lambda_{SW}$  with respect to the Coulomb moduli $s_i$
provides the basis of holomorphic differentials on the SW curve. 
We also note that the SW differential has a pole at $x=-m_a$.
The SW periods $\Pi ^{(0)}=(a_I^{(0)}, a_{DI}^{(0)})$ $(I=1,\cdots ,N_c-1)$  are defined by
\begin{align}
a^{(0)}_I=\oint _{\alpha _I} \lambda _{\text{SW}} , \qquad  a_{DI}^{(0)} =\oint _{\beta _I} \lambda _{\text{SW}} 
\end{align}
where $\alpha _I$ and $\beta _I$ are a pair of canonical one-cycles on the curve.
Here the superscript $(0)$ refers to the ``undeformed'' (or classical ) period.

On the Coulomb branch, there are singularities where some BPS particles become massless and the discriminant  of the SW curve (\ref{eq:swc}) becomes zero.
In particular, we are interested in  the superconformal point on the Coulomb branch where mutually non-local BPS particles become massless and the discriminant has higher order zero  \cite{Argyres:1995jj, Argyres:1995xn,Eguchi:1996vu,Gaiotto:2010jf}.

We consider the superconformal points of $SU(N_c)$ SQCD with $N_f$ flavors, where the SW curve degenerate maximally.
The related AD theories are classified in \cite{Eguchi:1996vu}.
We will discuss  the scaling limit around the superconformal point and the SW curves of the AD theories in detail.
\subsection{$N_f=0$}
For $N_f=0$,  the SW curve takes the form:
\begin{align}
    y^2=(x^{N_c}-s_2 x^{N_c-2}-\cdots -s_{N_c})^2-\Lambda^{2N_c}.
\end{align}
The superconformal point is given by $s_2=\cdots=s_{N_c-1}=0$ and $s_{N_c}=\pm \Lambda^{2N_c}$.  
Taking the limit $\epsilon\rightarrow 0$ with 
\begin{align}
    x=&\epsilon\Lambda\tilde{x}, \quad s_{i}=\pm\delta_{i, N_c}\Lambda^{N_c}+\epsilon^i \Lambda^i \tilde{u}_i \quad (i=2,\ldots, N_c),
\end{align}
the SW curve and the SW differential scale as
\begin{align}
    y^2=&\mp\epsilon^{N_c}\Lambda^{2N_c} 2\tilde{y}^2+\cdots ,\\
    \lambda_{SW}=&\pm\epsilon^{{N_c+3\over2}} \Lambda 2\sqrt{2}\tilde{\lambda}_{SW}+\cdots,
\end{align}
where
\begin{align}
    \tilde{y}^2=&\tilde{x}^{N_c}-\sum_{i=2}^{N_c} \tilde{u}_i \tilde{x}^{N_c-i}, \label{eq:swadnf0}
    \\
    \tilde{\lambda}_{SW}=&\tilde{y}d\tilde{x}. \label{eq:swdadnf0}
\end{align}
The curve (\ref{eq:swadnf0}) and $\tilde{\lambda}_{SW}$ defines the SW curve and the SW differential of the AD theory
of $(A_1, A_{N_c-1})$-type. 
The scaling dimensions of the variables in the curve are determined such that the BPS spectrum has unit dimension:  
$[\tilde{\lambda}_{SW}]=1$.
Then the scaling dimensions of $\tilde{u}_i$ are given by
as $[\tilde{u}_i]=2i/(N_c+2)$ ($i=2,\ldots, N_c$).
\subsection{$N_f=1$} 

For $N_f=1$, it is convenient to rewrite  the SW curve (\ref{eq:swc}) by the shift of $x$ as
\begin{align}
    y^2=&(x^{N_c}-s_1 x^{N_c-1}-\cdots-s_{N_c})^2-\Lambda^{2N_c-1}(x+m),
    \label{eq:swcnf1}
\end{align}
such that the superconformal point is realized at $x=0$.
It is found that the curve degenerate maximally at 
$s_i=\Lambda^i s^*_{i}$ and $m=\Lambda m^*$, where
\begin{align}
s^*_{N_c}=&\left\{
\begin{array}{cc}
-\left(
{({1\over4})_n ({3\over4})_n\over 2 \times n! ({3\over2})_n}
\right)^{1\over 4n+1}, &  N_c=2n+1,\\
\left(
{({3\over4})_n ({5\over4})_n\over 8 (n+1)! ({3\over2})_n}
\right)^{1\over 4n+3},
& N_c=2n+2,
\end{array}
\right.
\end{align}and
\begin{align}
s^*_{N_c-k}=&-{(-1)^k (-{1\over2})_k \over k!}{1\over (s^*_{N_c})^{2k-1}} \quad (k\geq 1),\nonumber\\
m^*=&(s^*_{N_c})^2 .
\end{align}
Here we define $(a)_n\equiv a(a+1) \cdots (a+n-1)$ for $n\geq 1$ and $(a)_0\equiv 1$.
At the superconformal point, the SW curve (\ref{eq:swcnf1}) becomes
\begin{align}
y^2=x^{N_c+1}(x^{N_c-1}+\cdots+{2N_c-1\over N_c+1} {\Lambda^{N_c-1}\over s^*_{N_c}} ).
\label{eq:swcnf1b}
\end{align}
We now take the scaling limit of the curve (\ref{eq:swcnf1})  around the fixed point 
as 
\begin{align}
x=&\epsilon \Lambda \tilde{x},\nonumber\\
s_1=&\Lambda s_1^*,\nonumber\\
s_i=&\Lambda^i (s_i^*+\sum_{k=1}^{i} \epsilon^k \tilde{s}_{i}^{(k)}),\quad  i=2,\cdots N_c-1\nonumber\\
m=&\Lambda (m^*+ \sum_{a=1}^{N_c} \epsilon^{a} \tilde{m}_{a} ),
\nonumber\\
s_{N_c}=&\Lambda^{N_c}(s^*_{N_c}+\sum_{k=1}^{N_c+1}\epsilon^i \tilde{s}_{k}),
\end{align}
with $\epsilon\rightarrow 0$.
The deviation of the moduli and the mass parameter from the superconformal point are chosen such that in the right hand side of (\ref{eq:swcnf1}), the coefficient in $\tilde{x}^i$ scales as $\epsilon^{N_c+1-i}$.
For example, let us consider the constant term which is given by $s_{N_c}^2-\Lambda^{2N_c-1} m$.
 Expanding this term in $\epsilon$ and requiring that the coefficients of $\epsilon^i$ ($i=0,\ldots, N_c$) are zero, we find that $\tilde{m}_a$ are determined by $\tilde{s}_k$ $(k=1,\ldots, N_c$) and 
 $ s_{N_c}^2-\Lambda m=\Lambda^{2N_c} \epsilon^{N_c+1} 2 s_{N_c}^* \tilde{s}_{N_c+1}+\cdots$.
 Similarly, for $i=2,\ldots, N_c-1$, $\tilde{s}_i^{(k)}$ can be determined by $\tilde{s}_1,\ldots, \tilde{s}_i$ by requiring that  the 
 coefficient in $\tilde{x}^{i}$ scales as $\epsilon^{N_c+1-i}$.
 Finally $\tilde{s}_1$ is fixed by $s^*_1$.
Then the SW curve scales as
\begin{align}
    y^2=&\Lambda^{2N_c} \epsilon^{N_c+1} {2N_c-1\over N_c+1}{1\over s^*_{N_c}}\tilde{y}^2+\cdots,
\end{align}
where 
\begin{align}
\tilde{y}^2=&\tilde{x}^{N_c+1}-\tilde{u}_2 \tilde{x}^{N_c-1}-\cdots- \tilde{u}_{N_c+1},
 \label{eq:adswnf1a}
\end{align}
and $\tilde{u}_2, \ldots, \tilde{u}_{N_c}$ are homogeneous functions of $\tilde{s}_2,\ldots, \tilde{s}_{N_c+1}$.
The SW differential becomes
    \begin{align}
    \lambda_{SW}=&\epsilon^{N_c+3\over2} {2\over s^*_{N_c} }\tilde{\lambda}_{SW}+\cdots,
\end{align}
where
\begin{align}
 \tilde{\lambda}_{SW}=&\tilde{y} d\tilde{x}.
 \label{eq:adswdnf1a}
\end{align}
The AD theory defined by (\ref{eq:adswnf1a}) and (\ref{eq:adswdnf1a}) is of the $(A_1,A_{N_c})$-type. The parameter $\tilde{u}_{i}$ has the scaling dimensions $[\tilde{u}_i]={2i \over N_c+3}$ $(i=2,\ldots, N_c+1)$.
\subsection{$N_f=2n+1$ ($n\geq 1$)}

For $N_f=2n+1$ ($n\geq 1$),  we redefine the moduli parameters of the SW curve (\ref{eq:swc}) by the shift of $x$:
\begin{align}
y^2=&(x^{N_c}-s_1 x^{N_c-1}-\cdots -s_{N_c})^2-\Lambda^{2N_c-N_f}\prod_{a=1}^{N_f}(x+m_a).
\label{eq:swnfodd1}
\end{align}
To find the superconformal point, we first set $m_a=m$ and introduce 
$\hat{x}=x+m$ such that the curve takes the form
\begin{align}
y^2=& (\hat{x}^{N_c}+u_1 \hat{x}^{N_c-1}+\cdots+u_{N_c})^2-\Lambda^{2N_c-N_f}\hat{x}^{N_f}.
\label{eq:swnfodd2}
\end{align}
Then the superconformal fixed point is realized at 
\begin{align}
u_{N_c}=u_{N_c-1}=\cdots=u_{N_c-n-2}=0,
\end{align}
where the SW curve degenerates into
\begin{align}
y^2
=&\hat{x}^{2n+1} \left\{ -\Lambda^{2N_c-2n-1} +\hat{x} (\hat{x}^{N_c-n-1}+u_1 \hat{x}^{N_c-n-2}+\cdots +u_{N_c-n-1})^2\right\}.
\end{align}
We  consider the scaling limit of the SW curve around the superconformal point.
First we introduce the scaling variables 
\begin{align}
x=&-m-\epsilon \Lambda \tm +\ep^2 \Lambda\tilde{x}, \\
m_a=&m+\ep \Lambda \tm +\ep^2 \Lambda \tc_a, \quad \sum_{a=1}^{N_f}\tc_a=0.
\end{align}
Then the SW curve (\ref{eq:swnfodd1}) becomes
\begin{align}
y^2=& (( -\ep \Lambda  \tm +\ep^2 \Lambda \tilde{x})^{N_c} +u_1 (-\ep  \Lambda \tm +\ep^2 \Lambda \tilde{x})^{N_c-1}+\cdots +u_{N_c})^2
-\Lambda^{2N_c} \ep^{2N_f} \prod_{a=1}^{N_f}(\tilde{x} +\tc_a).
\end{align}
We expand the first term in $\tx$:
 \begin{align}
 ( -\ep \Lambda \tm +\ep^2 \Lambda \tilde{x})^{N_c} +u_1 (-\ep \Lambda \tm +\ep^2\Lambda  \tilde{x})^{N_c-1}+\cdots +u_{N_c}
 =&(\ep ^2\Lambda \tilde{x})^{N_c} +\hat{u}_1 (\ep ^2\Lambda\tilde{x})^{N_c-1}+\cdots +\hat{u}_{N_c}.
 \end{align}
Then we rescale $\hat{u}_i$ as
\begin{align}
\hat{u}_{N_c-i}=&\ep^{N_f-2i}\Lambda^{N_c-i}\tu_{n -i}, \quad i=0,1,\cdots, n,
\end{align}
and $\hat{u}_i=O(1)$ ($i=1,\ldots, N_c-n+1$).
In the scaling limit $\epsilon\rightarrow 0$, the SW curve becomes
\begin{align}
y^2=&\ep^{2N_f} \Lambda^{2N_c} \tilde{y}^2+\cdots,
\end{align}
where
\begin{align}
\ty^2=& \tilde{C}(\tilde{x})^2-\tilde{G}(\tilde{x}),\label{eq:adswcnfodd1} \\
\tC(\tilde{x})=&\sum_{i=0}^{n}\tu_i \tilde{x}^{n-i}, \quad
\tG(\tilde{x})= \prod_{a=1}^{2n+1} (\tilde{x}+\tc_a).
\label{eq:asswcnfodd1a}
\end{align}
The SW differential  becomes
\begin{align}
\lambda_{SW}
=&-\ep^2\Lambda \tilde{\lambda}_{SW}+\cdots,
\end{align}
where
\begin{align}
\tilde{\lambda}_{SW}=&-\tilde{x} d\log {\tC-\ty \over \tC+\ty} .
\label{eq:adswdiffnfodd}
\end{align}
The curve (\ref{eq:adswcnfodd1}) and the differential (\ref{eq:adswdiffnfodd})
defines the AD theory for $N_f=2n+1$ ($n\geq 1$) theory. Note that the SW curve is that of $U(n)$ gauge theory with $N_f=2n+1$ hypermultiplets, which has $SU(2n+1)$ flavor symmetry.
The scaling dimensions of $\tilde{u}_i$ and $\tilde{c}_a$ are 
given by $[\tilde{u}_i]={2i+1\over2}$ and $[\tilde{c}_a]=1$.
Here $\tilde{u}_0$ and $\tilde{u}_1$ are relevant operators and $\tilde{u}_i$ for $i\geq 2$ are irrelevant operators.

\subsection{$N_f=2$}
For $N_f=2$, the superconformal fixed points of the theory are
given by
\begin{align}
s_i=&\pm \delta_{i, N_c-1}\Lambda^{N_c-1}, \quad m_a=0.
\label{eq:nf2fp}
\end{align}
The SW curve (\ref{eq:swc}) at this point becomes
\begin{align}
y^2=&x^{N_c+1}(x^{N_c-1}\mp 2\Lambda^{N_c-1}).
\end{align}
We take the scaling limit $\epsilon\rightarrow 0$ with
\begin{align}
x=&\ep^2 \Lambda \tilde{x},\nonumber \\
s_i =&\ep^{2i} \Lambda^i  \ts_i,\quad (i=2,\cdots N_c-2), \nonumber\\
s_{N_c-1}=&\pm \Lambda^{N_c-1} +\ep^{2N_c-2}\Lambda^{N_c-1}\ts_{N_c-1},\nonumber\\
s_{N_c}=&\ep^2 \tm  \Lambda^{N_c}+\ep^{2N_c} \Lambda^{N_c}\ts_{N_c},\nonumber\\
m_a=&\ep^2 \tm\Lambda+\ep^{N_c+1} \Lambda\tc_a, \quad (a=1,2)
\end{align}
where $\tc_1+\tc_2=0$.
In this limit, the SW curve (\ref{eq:swc}) scale as
\begin{align}
y^2
=&\mp 2 \ep^{2N_c+2}\Lambda^{2N_c}\ty^2+\cdots,
\end{align}
where
\begin{align}
\ty^2=&(\tilde{x}+\tm)\tilde{C}(\tilde{x})+{1\over2}\tc_1 \tc_2,
\quad \tC(\tilde{x})=\tilde{x}^{N_c}-\sum_{i=2}^{N_c}\ts_i \tilde{x}^{N_c-i}.
\end{align}
The SW differential is expanded as
\begin{align}
    \lambda_{SW}=&
     \mp 2\sqrt{-2} \ep^{N_c}\Lambda \tilde{\lambda}_{SW} +\cdots,
\end{align}
where
\begin{align}
\tilde{\lambda}_{SW}=&\ty d\log (\tilde{x}+\tm).
\end{align}
By the shift $\tilde{x}\rightarrow \tilde{x}-\tm$, we obtain 
\begin{align}
\ty^2=&\tilde{x}^{N_c+1}+\tu_1 \tilde{x}^{N_c}+\cdots+\tu_{N_c+1}, \label{eq:adswcnf2}\\
\tilde{\lambda}_{SW}=&\ty d\log \tilde{x}. \label{eq:adswdnf2}
\end{align}
The AD theory is of the type $(A_1, D_{N_c+1})$ and have $SU(2)$ flavor symmetry for $N_c$ odd and $U(1)$ for $N_c$ even.
Scaling dimensions of the moduli are given by $[\tu_i]={2i \over N_c+1}$ ($i=1,\cdots, N_c+1$).

\subsection{$N_f=2n$ ($n>1$)}
For $N_f=2n$ ($n>1$), we shift $x \rightarrow x-s_1/N_c$ and rewrite 
the SW curve as
\begin{align}
    y^2=&(x^{N_c}-s_1 x^{N_c-1}-\cdots -s_{N_c})^2-\Lambda^{2N_c-2n}
    \prod_{a=1}^{2n}(x+\hat{m}_a),
    \label{eq:swcnfe2}
\end{align}
where we choose $s_1$ such that $\sum_{a=1}^{2n}\hat{m}_a=0$.
The superconformal point is given by
\begin{align}
s_1=&s_2=\cdots =s_{N_c-n-1}=s_{N_c-n+1}=\cdots=s_{N_c}=0, \quad
s_{N_c-n}=\pm \Lambda^{N_c-n},\\
\hat{m}_a=&0.
\end{align}
At this point, the SW curve  (\ref{eq:swcnfe2}) becomes
\begin{align}
y^2
=&x^{N_c+n} (x^{N_c-n}\mp 2 \Lambda^{N_c-n}).
\end{align}
The scaling limit of this theory, which generalizes the $N_f=2$ case, has been discussed in \cite{Eguchi:1996vu}. 
However, it was pointed in \cite{Gaiotto:2010jf} that there exists another scaling limit which is consistent with the Cardy's a-theorem \cite{Cardy:1988cwa}, which was called the type A scaling, while the other is called the type B scaling.
We will discuss both type AD theories and their quantization.
\paragraph{type B scaling}
We begin with the type B scaling, in which we define the scaling variables 
\begin{align}
x=&\ep_B^2 \Lambda \tilde{x}, \quad 
\hat{m}_a=\ep_B^{N_c+2-n}\Lambda \tc_a, \quad 
s_i=\pm \Lambda^{N_c-n} \delta_{i, N_c-n}+\ep_B^{2i} \Lambda^{2i} \tilde{s}_i,
\label{eq:scalingnfeb}
\end{align}
where $\sum_{a=1}^{2n}\tilde{c}_a=0$.
The SW curve becomes
\begin{align}
y^2
=&\mp 2 \epsilon_B^{2n+2N_c} \Lambda^{2N_c}  \tilde{y}^2+\cdots,
\label{eq:scalingnfeby}
\end{align}
where 
\begin{align}
\tilde{y}^2=&\tilde{x}^{n} \tilde{C}(\tilde{x})\pm {1\over2}\tilde{x}^{2n-2} \tilde{C}_2, \label{eq:typeBSWc1}\\
\tilde{C}(\tilde{x})=&\tilde{x}^{N_c}-\sum_{i=2}^{N_c} \tilde{s}_{i}\tilde{x}^{N_c-i}.
\end{align}
We can express the curves as 
\begin{align}
\tilde{y}^2=&\tilde{x}^{N_c+n} +\sum_{i=1}^{N_c} \tilde{u}_{i}\tilde{x}^{N_c+n-i}.
\label{eq:adswcnfeb}
\end{align}
The SW differential becomes
\begin{align}
\lambda_{SW}=&\pm2\sqrt{-2} \Lambda\epsilon_B^2 \tilde{\lambda}_{SW}+\cdots,
\end{align}
where
\begin{align}
\tilde{\lambda}_{SW}=&\tilde{y}{d\tilde{x} \over \tilde{x}^n}.
\label{eq:adswdnfeb}
\end{align}
The scaling dimension of $\tilde{u}_i$ ($i=1,\ldots,N_c$) is ${2i\over N_c-n+2}$, where the operators for $i=N_c-n+3,\ldots, N_c$ are irrelevant.
The AD theory has $SU(2)$ flavor symmetry for $N_c-n+2$ even and $U(1)$ for $N_c-n+2$ odd.
\paragraph{Type A scaling}
Next we consider the type A scaling, in which we introduce the scaling parameters by
\begin{align}
x=&\ep_A^2 \Lambda \tilde{x},\\
\hat{m}_a=&\ep_A^2 \Lambda  \tc_a, \qquad \sum_{a=1}^{2n}\tc_a=0,
\\
s_{i}=&\pm \Lambda^{N_c-n} \delta_{i, N_c-n}+\ep_A^{-2N_c+2n+2i} \Lambda^{i} \tu_{i-N_c+n}, \quad
i=N_c-n,\ldots, N_c,
\end{align}
while  $s_1,\ldots, s_{N_c-n+1}$ are of order $O(1)$.
$\epsilon_A$ and $\epsilon_B$ are related by $\epsilon_A^2=\epsilon_B^{N_c-n+2}$ \cite{Gaiotto:2010jf}.
Then the SW curve scales as
\begin{align}
y^2=&\mp\ep_A^{4n} \Lambda^{2N_c} \ty^2+\cdots,
\end{align}
where
\begin{align}
\ty^2=& \tilde{C}(\tilde{x})^2-\tilde{G}(\tilde{x}), \label{eq:adswcnfea1}\\
\tC(\tilde{x})=&\sum_{i=0}^{n}\tu_i \tilde{x}^{n-i},
\quad
\tG(\tilde{x})=q\prod_{a=1}^{N_f}(\tilde{x}+\tc_a), \label{eq:adswcnfea1b}
\end{align}
with $\tu_0=1$ and $q=1$.
The SW differential scale as $\lambda_{SW}=\pm\epsilon_A^{2}\Lambda \tilde{\lambda}_{SW}+\cdots$, where
\begin{align}
\tilde{\lambda}_{SW}=&\tilde{x} d\log \left(
{\tC-\ty\over \tC+\ty}
\right). \label{eq:adswdnfea1}
\end{align}
The AD theory is ${\cal N}=2$ $U(n)$ gauge theory with $SU(2n)$ hypermultiplets
at strong coupling with $q=1$, where $q$ is the UV coupling.
The scaling dimensions are  $[\tu_i]=i$ ($i=1,\cdots, n$), $[\tc_a]=1$ $(a=1,\cdots, 2n$).

In the next section, we will study the quantum SW curves for the AD theories
which are obtained from the scaling limit of $SU(N_c)$ SQCDs.

\section{Quantum SW curves}
In this section, we study the quantum SW curves for the AD theories 
obtained from the scaling limit of ${\cal N}=2$ SQCD, where the theories have the SW curve of the form $\tilde{y}^2=f(\tilde{x})$ with the SW differential
$\tilde{\lambda}_{SW}$.
In the $(\tilde{x},\tilde{y})$ space,  the symplectic 
structure is introduced by the SW differential $d \tilde{\lambda}_{SW}$.
We then quantize the Poisson structure induced by the symplectic structure,
where the Planck constant corresponds to the deformation parameter 
in the Nekrasov-Shatashvili limit of the Omega-background.

\subsection{quantum SW curve for $SU(N_c)$ SQCD}
First we discuss the quantum SW curve for ${\cal N}=2$ $SU(N_c)$ SQCD \cite{Mironov:2009dv, Zenkevich:2011zx}.
It is convenient to write 
the SW curve (\ref{eq:swc}) and the SW differential (\ref{swdiff}) in the form
\begin{align}
C(x)-\frac{1}{2}\left( z+ \frac{G(x)}{z} \right) =0, \label{eq:swc2} \\
\lambda _{\text{SW}} = x \left( d \log G(x) -2 d\log z \right),
\end{align}
by introducing
\begin{align}
z=y+C(x).
\end{align}
The SW differential defines the holomorphic symplectic form $d\lambda _{\text{SW}} =-2 d x\wedge d \log z$.
Now we quantize the SW curve by replacing $z$ by the differential operator 
\begin{align}
\hat{z}=\exp \left( -i \hbar \frac{\partial }{\partial x} \right),
\end{align}
and define the quantum SW curve by
\begin{align}
\left[ \frac{1}{2} \left( \exp \left( -i \hbar \frac{\partial }{\partial x}\right) +\exp \left( i \frac{ \hbar}{2} \frac{\partial }{\partial x}\right) G(x) \exp \left( i \frac{ \hbar}{2} \frac{\partial }{\partial x}\right) \right) -C(x) \right] \Psi (x)=0.
\label{eq:qswcsqcd1}
\end{align}
Here we take the ordering prescription of the differential operators as in \cite{Zenkevich:2011zx} such that the corrections to the WKB periods are even power series in $\hbar$.

\subsection{Quantum SW curves of AD theories}
We now study the quantum SW curves for the AD theories obtained by the scaling limit of ${\cal N}=2$ $SU(N_c)$ SQCD discussed in the previous section.
They are classified into the Schr\"odinger type and the SQCD type.
The Schr\"odinger type quantum SW curve is the second order differential equations with  a potential term, while the SQCD type quantum SW curve is the
differential equation of the form (\ref{eq:qswcsqcd1}) with some
$C(x)$ and $G(x)$.

\subsubsection{the Schr\"odinger type quantum SW curve}
For $N_f=0,1,2$ and $N_f=2n$ ($n>1$) with type B scaling, the quantum SW curve
corresponds to the Schr\"odinger equation defined in the complex plane.

\paragraph{\underline{$N_f=0,1$}}
For $N_f=0$ and $N_f=1$, the SW curve is given by (\ref{eq:swadnf0}) and (\ref{eq:adswnf1a}), respectively.
The SW curve is written in the form
\begin{align}
\tilde{y}^2=&-Q(\tilde{x}), \quad
Q(\tilde{x})=-\tilde{x}^{N}+\sum_{i=2}^{N}\tilde{u}_i \tilde{x}^{N-i}
\label{eq:qswnf01}
\end{align}
where $N=N_c$ for $N_f=0$ and $N=N_c+1$ for $N_f=1$. 
The SW
differential is given by $\tilde{\lambda}_{SW}=\tilde{y}d\tilde{x}$, which introduces the symplectic form $d\tilde{\lambda}_{SW}=d\tilde{y}\wedge d\tilde{x}$. 
The quantum SW curve is obtained by $\tilde{y}\rightarrow -i\hbar {\partial \over \partial \tilde{x}}$:
\begin{align} \label{qswc:class0}
\left( -\hbar ^2 \frac{\partial ^2}{\partial \tilde{x}^2 } +Q(\tilde{x}) \right) \Psi (\tilde{x}) =0.
\end{align}

\paragraph{\underline{$N_f=2$}}
For $N_f=2$, the SW curve and the SW differential are  (\ref{eq:adswcnf2}) and  (\ref{eq:adswdnf2}).
Introducing $\xi=\log \tx$, the SW curve is
written in the form
\begin{align}
\tilde{y}^2 =&-Q(\xi),\quad
Q(\xi)=-\sum _{i=0}^{N_c+1} \tilde{u}_i e^{(N_c+1-i) \xi },
\end{align}
with $\tilde{u}_0=1$.
The SW differential becomes $\tilde{\lambda}_{SW}=\tilde{y}d\tilde{\xi}$.
Then
the quantum SW curve is given by
\begin{align} \label{qswc:class4_2}
\left( -\hbar ^2 \frac{\partial ^2}{\partial \xi ^2} +Q(
\xi ) \right) \Psi (\xi )=0. 
\end{align}

\paragraph{\underline{$N_f=2n$ ($n>1$) with type B scaling}}
For $N_f=2n$ with type B scaling, the SW curve and the SW differential are
given by (\ref{eq:adswcnfeb}) and (\ref{eq:adswdnfeb}), respectively.
Introducing $\xi=\tilde{x}^{1-n}$, they become
\begin{align}
\tilde{y}^2=&-Q(\xi),\quad Q(\xi)=-\sum_{i=0}^{N_c} \tilde{u}_i \xi^{N_c-n+1\over 1-n},
\label{eq:adswpotnfeb}\\
\tilde{\lambda}_{SW}=&{1\over 1-n} \tilde{y} d\xi.
\end{align}
We then find the quantum SW curve (\ref{qswc:class4_2}) with 
the potential $Q(\xi)$ defined by (\ref{eq:adswpotnfeb}).
In the next section, we will compute the quantum periods and their relation to the classical periods.
We will compare the quantum periods of the AD theory 
with the scaling limit of the periods of the UV theory.
It will be shown that we need to include the quantum correction to the potential.

\subsubsection{SQCD type quantum SW curve}

\paragraph{\underline{$N_f=2n+1$ ($n\geq 1$)}}
For $N_f=2n+1$ ($n\geq 1$) case, the SW curve and the SW differential are given by (\ref{eq:adswcnfodd1}) and (\ref{eq:adswdiffnfodd}), respectively.
This is the curve of the SQCD type for non-asymptotic free theory.
Then the quantum SW curve is given by
\begin{align}
\left[ \frac{1}{2} \left( \exp \left( -i \hbar \frac{\partial }{\partial \tilde{x}}\right) +\exp \left( i \frac{ \hbar}{2} \frac{\partial }{\partial \tilde{x}}\right) \tilde{G}(\tilde{x}) \exp \left( i \frac{ \hbar}{2} \frac{\partial }{\partial \tilde{x}}\right) \right) -\tilde{C}(\tilde{x}) \right] \Psi (\tilde{x})=0,
\label{eq:qswcsqcd2}
\end{align}
where $\tilde{C}(\tilde{x})$ and $\tilde{G}(\tilde{x})$ are given by
(\ref{eq:asswcnfodd1a}).

\paragraph{\underline{$N_f=2n$ $(n\geq 2)$ with type A scaling}}
For $N_f=2n$ ($n>1$) with type A scaling,
the SW curve and the SW differential is given by (\ref{eq:adswcnfea1}) and (\ref{eq:adswdnfea1}).
Then the quantum SW curve takes the form of (\ref{eq:qswcsqcd2}) with 
$\tilde{C}(\tilde{x})$ and $\tilde{G}(\tilde{x})$
which are defined by 
(\ref{eq:adswcnfea1b}).

\section{Quantum SW periods}
In the previous section,
we have written down the quantum SW curves  for the AD theories obtained from the scaling limit of
 ${\cal N}=2$ SQCD.
In the section, we will study the WKB expansions of the quantum SW curve and the quantum SW periods.
In particular, we will investigate the relation between the classical periods and their quantum corrections. 
They are shown to be related by the differential operators with 
respect to the moduli parameters.
These relations are important to study the quantum periods as a section of the Coulomb branch 
because their relations are valid  in any locus of the moduli space of AD theories.
In the present work, we will write down the quantum corrections up to the 
fourth order in the Planck constant.
We will also check whether the quantum corrections to the quantum periods
of the AD theories are consistent with the scaling limit of the 
those of ${\cal N}=2$ SQCD.

\subsection{WKB solution to quantum SW curve}
We study the quantum SW periods for the AD theories based on the WKB solution to the quantum SW curve.
We first consider the WKB solution of the SQCD type quantum SW curve. 
This type of differential equation
includes  ${\cal N}=2$ SQCD,  the AD theory associated with $N_f=2n+1$ ($n\geq 1$) and the
type A scaling of $N_f=2n$ ($n>1$) theory.

Substituting
the WKB solution of the form
\begin{align}
\Psi(x)=&\exp\left( {i\over \hbar}\int^x dx' P(x')\right), \quad P(x)=\sum_{n=0}^{\infty}\hbar^n p_n(x),
\label{eq:wkb1}
\end{align}
into (\ref{eq:qswcsqcd1}), we can determine $p_n(x)$ recursively.  
We define the quantum SW period (the WKB period) by
\begin{align}
\Pi_\gamma=&\int_\gamma P(x) dx,
\end{align}
where $\gamma$ is a 1-cycle on the SW curve.
The quantum SW period is expanded as
\begin{align}
    \Pi_\gamma=&\sum_{n=0}^{\infty} \hbar^n \Pi_\gamma^{(n)},\quad 
    \Pi_\gamma^{(n)}=\int_\gamma p_n(x)dx.
    \label{eq:wkb43}
\end{align}
For odd $n$, $p_n(x)$  becomes the total derivative in $x$, which does not contribute to the WKB periods. 
For even $n=0$,  $p_0(x)=\log(C(x)+y)$.  
For $n=2,4$, up to total derivative terms, $p_2(x)$ and $p_4(x)$ are found to be \cite{Ito:2019twh}:
\begin{align}
    p_2(x)=&{C''\over 8y}-{C (y')^2\over 8y^3},\\
    p_4(x)=&{(C')^2 C'' \over 768 y^3}
-{7\over 1536} {C (C'')^2\over y^3}
-{C (C')^2 y''\over 256 y^4}
\nonumber\\
&+{11\over 512}{ C^2 C'' y''\over y^4}
-{11\over 1536} {C'' y''\over y^2}
-{5 C^3 (y'')^2\over 256 y^5}
+{17 C (y'')^2\over 1536 y^3}
\nonumber\\
&-{C C' C^{(3)} \over 384 y^3}
+{C^2 C' y^{(3)}\over 128 y^4}
-{C' y^{(3)} \over 384 y^2}
\nonumber\\
&-{3 C^2 C^{(4)} \over 1024 y^3}
+{11 C^{(4)} \over 3072 y}
+{3 C^3 y^{(4)} \over 1024 y^4}
-{5 C y^{(4)} \over 1024 y^2}.
\end{align}

We next consider the WKB solutions to the Schr\"odinger type quantum SW
curve, which corresponds to the AD theories associated with $N_f=0,1,2$ and type B scaling of $N_f=2n$ ($n>1$) SQCD. 
The quantum SW curve takes the form
\begin{align}
    \left( -\hbar^2 {d^2\over dx^2}+Q(x)\right)\Psi(x)=&0,
    \label{eq:schro1}
\end{align}
where we consider the case that the potential includes the quantum correction
\begin{align}
    Q(x)=&Q_0(x)+\hbar^2 Q_2(x).
\end{align}
The WKB solution (\ref{eq:wkb1}) to (\ref{eq:schro1}) has been studied  (see for example \cite{Ito:2019twh}).
$p_n$ for odd $n$ is  a total derivative again and 
$p_n$ ($n=0,2,4$) are found to be
\begin{align} \label{recursion_relation}
  p_0(x)=&i \sqrt{Q_0},\\
  p_2(x)=&{i\over2}{Q_2\over \sqrt{Q_0}}+{i\over 48}{Q_0''\over Q_0^{3\over2}} ,\label{recursion_relation2}\\
  p_4(x)=&-{7i\over 1536}{Q''_0{}^2\over Q_0^{7\over2}}+{i\over 768}{Q_0^{(4)}\over Q_0^{5\over2}}
  -\frac{i}{32}{ Q_2 Q''_0\over  Q_0^{5\over2}}
  +\frac{i}{48}{ Q''_2\over  Q_0^{3\over2}}
  -\frac{i}{8}{ Q_2^2\over  Q_0^{3\over2}},
  \label{recursion_relation4}
\end{align} 
up to total derivatives.

\subsection{Quantum SW periods and the scaling limit}
We now study the quantum SW periods of the AD theories and compare them with the quantum SW periods of the SQCD, which has been studied for $SU(2)$ SQCD \cite{Ito:2018hwp}.  It  has  been shown that the quantum SW periods for the AD theories are also obtained from the scaling limit of the SQCD.
We consider the scaling limit of higher order terms in the WKB series in $SU(N_c)$ SQCD with $N_f$ flavors.
For $N_f=0$, in  \cite{Ito:2019twh}, the scaling limit of ${\cal N}=2$ $SU(N_c)$ super Yang-Mills theory has been shown to be  consistent with the quantum SW curve of the $(A_1,A_{N_c-1})$-type AD theory.
In the case of $SU(N_c)$ SQCD with $N_f$ flavors,
we confirm  that the scaling limit of $p_2$ and $p_4$ in $N_f=2,2n+1$ ($n\geq 0$)
and type A scaling of $N_f=2n$ ($n>1$) theory are consistent with the WKB solutions  of the corresponding AD theories.
For type B scaling of $N_f=2n$ ($n>1$) theory, however,  the scaling limit of the SQCD leads to the quantum correction to the potential term of the quantum SW curve of the AD theory. 
This is similar to the case of the scaling limit of ${\cal N}=2$ $SO(2N)$ super Yang-Mills theory \cite{Ito:2019twh} whose SW curve is obtained from the $SU(2N)$ with massless $N_f=4$ hypermultiplets \cite{Brandhuber:1995zp}.

Now we focus on  the type B scaling of $N_f=2n$ theory.
Let us consider the scaling limit of $p_2$.
In this limit, $C(x)$ scales as
\begin{align}
C(x)=&-\epsilon^{2n}\Lambda^{2N_c}\tilde{x}^n+\epsilon^{2N_c}\Lambda^{2N_c}\tilde{C}(\tilde{x}).
\end{align}
From this and (\ref{eq:scalingnfeby}), we find
\begin{align}
    p_2=&-{\epsilon^{n-N_c-4} \over \Lambda^{2}}
    \left\{ {1\over24} {n\tilde{x}^{n-1} Q' \over Q^{3/2}}-{1\over 48}{\tilde{x}^n Q'' \over Q^{3/2}}+\cdots
    \right\},
\end{align}
where $Q(\tilde{x})$ is the r.h.s. of eq. (\ref{eq:typeBSWc1}).
Comparing this formula with (\ref{recursion_relation2}), in which  $Q_0(\xi)$ given by (\ref{eq:adswpotnfeb}), we find that $Q_2(\xi)$ is given by
\begin{align}
    Q_2(\xi)=&{n\over 4(1-n)} {1\over \xi^2}.
    \label{eq:typebpot2}
\end{align}
We can also check the scaling limit of $p_4$ reduces to that of the AD theory
with the same $Q_2(\xi)$.
The existence of corrections to the potential depends on the flavor symmetry 
which are kept manifestly in the quantum SW curve \cite{Ito:2019twh}.

\subsection{Differential operator for higher order corrections}
We now study the corrections (\ref{eq:wkb43}) in the quantum SW periods by expressing them
by  the classical periods $\Pi^{(0)}_{\gamma}$:
\begin{align}
\Pi^{(2n)}_{\gamma}=&{\cal O}_{2n}\Pi^{(0)}_{\gamma},
\label{eq:pfop1}
\end{align}
where ${\cal O}_{2n}$ is some differential operator with respect to the moduli parameters. 
This relation is useful to evaluate the quantum periods explicitly from 
the classical periods.
The relation (\ref{eq:pfop1}) is obtained from the relation $p_{2n}={\cal O}_{2n}p_0$ up to total derivative terms. 
To distinguish the quantum WKB periods of AD theories and those of the SQCD, 
we use the notation $\tp_n$ instead of $p_n$
for the AD theories.

\subsubsection{$N_f=0,1$}
We begin with the WKB periods for the quantum SW curve of the Schr\"odinger type.
For the quantum SW curve (\ref{eq:qswnf01}) of the AD theories associated with $N_f=0,1$, the operators ${\cal O}_2$ and ${\cal O}_4$ have been constructed in \cite{Ito:2019twh}, where
$\tilde{p}_2$ and $\tilde{p}_4$ are given by
\begin{align}
 \tilde{p}_2
 =&-{1\over 24}
 \left(
  -[N]_2
   \partial_{\tu_2} \partial_{\tu_{N}}
 + \sum_{j=2}^{N-2} [N-j]_2 \tu_j \partial_{\tu_{j+2}} \partial_{\tu_{N}} 
 \right) \tilde{p}_0,
 \label{eq:nf01p2}
 \end{align}
\begin{align}
\tilde{p}_4
=&
\left(\frac{7}{5760}
 \Bigl\{ ([N]_2)^2
   \partial_{\tu_2}^2 
 -2 [N]_2 \sum_{j=2}^{N-3}
 [N-j]_2 \tu_j \partial_{\tu_2}\partial_{\tu_{j+2}}
 \right.
 \nonumber\\
&+
 \sum_{j=2}^{N-3}  \sum_{k=2}^{N-3}
 [N-k]_2 [N-j]_2
 \tu_j \tu_k
 \partial_{\tu_{j+2}}\partial_{\tu_{k+2}}
 \Bigr\}
 \partial_{\tu_{N}}^2
 \notag \\
&\left.+\frac{1}{1152}
 \Big\{ - [N]_4
\partial_{\tu_4}
+\sum_{j=2}^{N-4}
[N-j]_4 
\tu_j  \partial_{\tu_{j+4}}
\Bigr\}  \partial_{\tu_{N}}^2
\right) \tilde{p}_0.
\label{eq:nf01p4}
\end{align}
where we define $[a]_k:=a(a-1) \cdots (a-k+1)$ for $k\geq 1$ and $[a]_0:=1$.
Here $N=N_c$ for $N_f=0$ and $N=N_c+1$ for $N_f=1$.
These formulas follow from 
\begin{align}
\pa_{\tu_{j_1}}\cdots \pa_{\tu_{j_n}}(-Q)^{1\over2}=&\left[ {1\over2} \right]_n{\pa_{\tu_{j_1}}(-Q) \cdots \pa_{\tu_{j_n}}(-Q)\over (-Q)^{n-1/2}},
\end{align} 
and
\begin{align}
\pa_{\tilde{x}}^k Q=&\sum_{i=0}^{N-k} [N-i]_k \tu_i  \pa_{\tu_{i+k}}Q,
\end{align}
where $Q(\tilde{x})$ is defined by (\ref{eq:qswnf01}).

\subsubsection{$N_f=2$}
For $N_f=2$, the quantum SW curve is given by (\ref{qswc:class4_2}).
Then $\tilde{p}_0=iQ^{1/2}$ and $\tilde{p}_2$ and $\tilde{p}_4$ are given by
(\ref{recursion_relation2}) and (\ref{recursion_relation4}).
From
\begin{align}
\partial_{\tilde{u}_{i_1}}\cdots \partial_{\tilde{u}_{i_k}}
\tilde{p}_0=&-i \left[{1\over2}\right]_k{\exp(\sum_{j=1}^{k} (N_c+1-i_j)) \over Q^{(2k-1)/2}},
\end{align}
we get
\begin{align}
\tilde{p}_2=&{1\over12} \sum_{i=0}^{N_c} \tilde{u}_i (N_c+1-i)^2 \partial_{\tilde{u}_{i+1}}\partial_{\tilde{u}_{N_c}}\tilde{p}_0,
\label{eq:nf2p2}
\end{align}
and
\begin{align}
\tp_4=& {7\over 1440} \sum_{i=0}^{N_c} \sum_{j=0}^{N_c}
\tilde{u}_i \tilde{u}_j (N_c+1-i)^2(N_c+1-j)^2 \partial_{\tilde{u}_{i+1}}\partial_{\tilde{u}_{j+1}}\partial_{\tilde{u}_{N_c}}^2 \tp_0
\nonumber\\
&-{1\over288} \sum_{i=0}^{N_c} \tilde{u}_i (N_c+1-i)^4 \partial_{\tilde{u}_{i+1}}\partial_{\tilde{u}_{N_c}}\partial_{\tilde{u}_{N_c+1}}\tp_0.
\label{eq:nf2p4}
\end{align}

\subsubsection{$N_f=2n$ ($n>1$) with type B scaling }
The quantum SW curve  is the Schr\"odinger type with the potential $Q(\xi)+\hbar^2 Q_2(\xi)$, where $Q(\xi)$ and $Q_2(\xi)$ are given by
(\ref{eq:adswpotnfeb}) and (\ref{eq:typebpot2}), respectively.
From 
\begin{align}
\partial_{\tilde{u}_{i_1}}\cdots \partial_{\tilde{u}_{i_k}}
\tilde{p}_0=&-i \left[{1\over2}\right]_k{\prod_{j=1}^{k} \xi^{(N_c+1-i_j )\over 1-n} \over Q^{(2k-1)/2}},
\end{align}
we obtain
\begin{align}
\tp_2=&{1\over12} \sum_{i=0}^{N_c}\tilde{u}_i 
[\mu(i)]_2
\partial_{\tilde{u}_{i+1}}\partial_{\tilde{u}_{N_c-n+1}}\tp_0
+{1\over12}\tilde{u}_{N_c} {n(2n-1)\over (1-n)^2} \partial_{\tilde{u}_{N_c}}\partial_{\tilde{u}_{N_c-n+2}}\tp_0-
A \partial_{\tilde{u}_{N_c-n+2}}\tp_0,
\label{eq:p2nfeventypeb}
\end{align}
and 
\begin{align}
\tp_4=&{7\over 1440}\Bigl\{
\sum_{i=0}^{N_c-1}\sum_{j=0}^{N_c-1}
\tilde{u}_i \tilde{u}_j [\mu(i)]_2 [\mu(j)]_2
\partial_{\tilde{u}_{i+1}}\partial_{\tilde{u}_{j+1}}\partial_{\tilde{u}_{N_c-n+1}}^2 \tp_0
\nonumber\\
&+2 \sum_{i=0}^{N_c-1}\tilde{u}_i \tilde{u}_{N_c} 
[\mu(i)]_2 [\mu(N_c)]_2 
 \partial_{\tilde{u}_{i+1}} \partial_{\tilde{u}_{N_c}} \partial_{\tilde{u}_{N_c-n+1}} \partial_{\tilde{u}_{N_c-n+2}} \tp_0
+\tilde{u}_{N_c}^2 ([\mu(N_c)]_2)^2
\partial_{\tilde{u}_{N_c}}^2 \partial_{\tilde{u}_{N_c-n+2}}^2 \tp_0
\Bigr\}
\nonumber\\
&
+{1\over 288} \Bigl\{ \sum_{i=0}^{N_c-1} \tilde{u}_i [\mu(i)]_4
\partial_{\tilde{u}_{i+1}}\partial_{\tilde{u}_{N_c-n+2}}\partial_{\tilde{u}_{N_c-n+1}}\tp_0
+\tilde{u}_{N_c} [\mu(N_c)]_4 
\partial_{\tilde{u}_{N_c}}\partial_{\tilde{u}_{N_c-n+2}}^2 \tp_0
\Bigr\}
\nonumber\\
&
+{A\over32} \Bigl\{ \sum_{i=0}^{N_c-1} \tilde{u}_i [\mu(i)]_2 
\partial_{\tilde{u}_{i+1}}\partial_{\tilde{u}_{N_c-n+2}}\partial_{\tilde{u}_{N_c-n+1}}\tp_0
+\tilde{u}_{N_c} [\mu(N_c)]_2 
\partial_{\tilde{u}_{N_c}}\partial_{\tilde{u}_{N_c-n+2}}^2 \tp_0
\Bigr\}
\nonumber\\
&-{A(1-A)\over 2}\partial_{\tilde{u}_{N_c-n+2}}^2\tp_0.
\label{eq:p4nfeventypeb}
\end{align}
where we defined
\begin{align}
    \mu(i):={N_c+n-i\over 1-n}, \quad A={n\over 4(1-n)}.
\end{align}

\subsubsection{$N_f=2n+1$ and $N_f=2n$ $(n>1)$ with type A scaling}
Finally we study the quantum corrections for the quantum SW curve of SQCD type, which is more complicated than the Schr\"odinger type.
For the AD theories $N_f=2n+1$ theories and the type A scaling of $N_f=2n$ SQCD, we can parametrize the curves as
\begin{align}
\ty^2=&\tC(\tx)^2-\tG(\tx), \label{eq:swsqcd1a}\\
\tC(\tx)=&\sum_{i=0}^{N}\tu_i \tx^{N-i}, \\
\tG(\tx)=&\sum_{a=0}^{N_f}\tv_a \tx^{N_f-a}, \label{eq:swsqcdg1a}
\end{align}
where 
 $N=N_c$, $N_f=2n+1$, $\tv_0=1$ and $\tv_1=0$ for the $N_f=2n+1$ theory,
 $N=2n$, $N_f=2n$, $\tu_0=1$ and $\tv_1=0$ for the $N_f=2n$ theory.
For $N_f=2n$ theory, however, as noticed in $SU(2)$ SQCD with $N_f=4$ \cite{Ito:2017iba}, we cannot find enough number of the basis of meromorphic
differentials on the SW curve by the derivatives of the SW differential.
In \cite{Ito:2017iba}, we use the parameter $q=\exp(2\pi i\tau_{UV})$, evaluated at the UV coupling, to obtain an additional basis of the differentials.  
Therefore we set $\tv_0=q$ in (\ref{eq:swsqcdg1a}) and consider the limit $q\rightarrow 1$ after the calculation of the quantum periods.
Instead,  rescaling $x$ and $y$, we can normalize $\tv_0=1$ and regard $\tu_0$ as the new moduli parameter of the curve. 
Then the SW curves for $N_f=2n+1$ and type A scaling of $N_f=2n$ can be described in the same manner by (\ref{eq:swsqcd1a}) with $\tv_0=1$ and $\tv_1=0$.
We now calculate the differential operators ${\cal O}_2$ and ${\cal O}_4$
for this parametrization of the curve. 

To calculate $\tp_2$, we first rewrite it as
\begin{align}
    \tp_2=&-{\tC^2 \partial_{\tx}^2 \tC\over 24 \ty^{3}}+{\partial_{\tx}^2 C \over 24 \ty} 
-{1\over48}{\partial_{\tx} \tC \partial_{\tx} \tG\over \ty^{3}}
+{\tC \partial_{\tx}^2 \tG\over 48 \ty^{3}},
\end{align}
up to total derivative terms.
Then using the formulas
\begin{align}
    \pa_{\tu_i}\tp_0=&{\pa_{\tu_i}\tC\over \ty},\\
     \pa_{\tu_i}\pa_{\tu_j}\tp_0=&-{\tC \pa_{\tu_i}\tC \pa_{\tu_j}\tC \over \ty^3},\\
     \pa_{\tu_i}\pa_{\tv_a}\tp_0&
={1\over2}{ \pa_{\tu_i}\tC \pa_{\tv_a}\tG \over \ty^3},
\end{align}
and
\begin{align}
    \pa^k_{\tx} \tC=&\sum_{i=0}^{N-k} \tu_i [N-i]_k \partial_{\tu_{i+k}}\tC, \quad k\geq 0\\
    \pa^k_{\tx} \tG=&\sum_{i=a}^{N_f-k} \tv_a [N_f-a]_k \partial_{\tv_{a+k}}\tG, \quad k\geq 1,
\end{align}
we express $\tp_2$ as
\begin{align}
\tp_2=&{1\over24} 
\left(
\sum_{i=0}^{N} \sum_{j=0}^{N-2} \tu_i \tu_j [N-j]_2 
\partial_{\tu_i}  \partial_{\tu_{j+2}}
+\sum_{j=0}^{N-2} \tu_j [N-j]_2  
\partial_{\tu_{j+2}}
\right)\tp_0
\nonumber\\
&
-{1\over24} \sum_{i=0}^{N-1}  \tu_i (N-i) \left(
N_f \partial_{\tu_i} \partial_{\tv_2} +\sum_{a=2}^{N_f-1} \tv_a (N_f-a) \partial_{\tu_{i+1}}\partial_{\tv_{a+1}} 
\right)\tp_0
\nonumber\\
&
+{1\over24}\sum_{i=0}^{N} \sum_{a=0}^{N_f-2}  \tu_i \tv_a 
[N_f-a]_2 
 \partial_{\tu_i} \partial_{\tv_{a+2}}
\tp_0.
\label{eq:p2_sqcd}
\end{align}
Next we study $\tp_4$, which is given by
\begin{align}
    \tp_4=&\pa^4_{\tx} \tG \left( {19\over 11520} {\tC\over \ty^{3}}-{41\over 23040}{\tC^3\over \ty^{5}}\right)
+\pa_{\tx}^3 \tG \left(
{11\over 4608} {\tC^4 \pa_{\tx} \tC\over \ty^{7}}-{29\over 7680}{\tC^2 \pa_{\tx} \tC\over \ty^{5}}
+{1\over 720}{\pa_{\tx} \tC\over \ty^{3}}
\right)
\nonumber\\
&+\pa_{\tx}^2 \tG \left(
{7\over 768} {\tC^3 (\pa_{\tx} \tC)^2\over \ty^{7}}
-{7\over 768} {\tC (\pa_{\tx} \tC)^2\over \ty^{5}}
+{7 \over 384} {\tC^4\pa_{\tx}^4 \tC\over \ty^{7}}
-{17\over 768} {\tC^2 \pa_{\tx}^2 \tC\over \ty^{5}}
+{19\over 3840} {\pa_{\tx}^2 \tC\over \ty^{3}}
\right)
\nonumber\\
&-{11\over 9216} {\pa_{\tx}^3 \tG \pa_{\tx} \tG  \tC^3\over \ty^{7}}
+{7\over 1536} {\pa_{\tx}^2 \tG\pa_{\tx} \tG \tC^2 \pa_{\tx} \tC \over \ty^{7}}
-{7\over 384}{\tC^5 (\pa_{\tx}^2 \tC)^2\over \ty^{7}}
+{29\over 960} {\tC^3 (\pa_{\tx}^2 \tC)^2\over \ty^{5}}
\nonumber\\
&-{23\over 1920} {\tC (\pa_{\tx}^2 \tC)^2\over \ty^{3}}
+{1\over 384} {\tC^4 \pa_{\tx}^4 \tC \over \ty^{5}}
-{11\over 2880} {\tC^2 \pa_{\tx}^4 \tC\over \ty^{3}}
+{7\over 5760} {\pa_{\tx}^4 \tC\over \ty^{}}.
\label{eq:p4_sqcd}
\end{align}
Here we need to express ${\tC^{n_0} \pa_{\tx}^{n_1} \tC \cdots \pa_{\tx}^{m_1} \tG\cdots \over \ty^{2l+1}}$ in terms of the derivatives of $\tp_0$.
In order to simplify the formulas, it is convenient to introduce the differential operator defined by
\begin{align}
{\cal D}_{(n_1 n_2 \cdots | m_1 m_2\cdots)}&:=\sum_{i_1,i_2,\cdots=0}^{N_c}
\sum_{a_1,a_2,\cdots=0}^{N_f} \tu_{i_1} [N_c-i_1]_{n_1} \tu_{i_2} [N_c-i_2]_{n_2}\cdots
\tv_{a_1} [N_f-a_1]_{m_1} \tv_{a_2} [N_f-a_2]_{m_2}\cdots
\nonumber\\
& \pa_{\tu_{i_1+n_1}}\pa_{\tu_{i_2+n_2}}\cdots \pa_{\tv_{a_1+m_1}}\pa_{\tv_{a_2+m_2}}\cdots .
\label{eq:diffop_sqcd}
\end{align}
Then we find the relations such as
\begin{align}
    {\pa_{\tx}^4 \tG \tC \over \ty^3}=&2{\cal D}_{(0|4)}\tp_0,\quad
{\tC^3 \pa_{\tx}^4 \tG \over \ty^5}=-{2\over3} {\cal D}_{(00|4)}\tp_0.
\end{align}
Then the first terms of (\ref{eq:p4_sqcd}) can be written in terms of $\tp_0$.
Other terms in (\ref{eq:p4_sqcd}) can also be written by using (\ref{eq:diffop_sqcd}). 
In appendix A, we present a set of identities for $\tp_4$.
Finally we find that
\begin{align}
{\cal O}_4
=&{19\over 5760}{\cal D}_{(0|4)}+{41 \over 34560}{\cal D}_{(00|4)}
+{11\over 34560} {\cal D}_{(001|3)}
+{19\over 8640}  {\cal D}_{(01|3)}+{1\over 360} {\cal D}_{(1|3)}
\nonumber\\
&+{7\over 5760}  {\cal D}_{(011|2)}
+{7\over 1440}  {\cal D}_{(11|2)}
+{7\over 2880} {\cal D}_{(004|2)}
+{71\over 5760}  {\cal D}_{(04|2)}+{19\over 1920}  {\cal D}_{(2|2)}
\nonumber\\
&
+{11\over 34560} 
{\cal D}_{(00|13)}
-{7\over 5760} {\cal D}_{(01|12)}
+{7\over 5760} {\cal D}_{(0022|)}
\nonumber\\
&+{37\over 5760 }  {\cal D}_{(022|)}
+{1\over 180} {\cal D}_{(22|)}
+{1\over 1152} {\cal D}_{(004|)}
-{17\over 5760}{\cal D}_{(04|)}
+{7\over 5760}{\cal D}_{(4|)}.
\label{eq:diffop_sqcd_p4}
\end{align}
We note that in (\ref{eq:diffop_sqcd_p4})  the operators ${\cal D}_{(00|13)}$ and ${\cal D}_{(00|12)}$ appear, where in their formal definition (\ref{eq:diffop_sqcd}) , there exist differential operator with respect to $\tv_1$.
Since the parameter $\tv_1$ does not exist in the curve, we redefine these derivative operators as
\begin{align}
{\cal D}_{(00|13)}
=&\sum_{i_1=0}^{N_c} \sum_{i_2=0}^{N_c} \sum_{a_1=1}^{N_f-1} \sum_{a_2=0}^{N_f-3}
\tu_{i_1} \tu_{i_2} \tv_{a_1} (N_f-a_1)\tv_{a_2}[N_f-a_2]_3 \pa_{\tu_{i_1}}\pa_{\tu_{i_2}}\pa_{\tv_{a_1+1}}\pa_{\tv_{a_2+3}}
\nonumber\\
&+\sum_{i_1=0}^{N_c} \sum_{i_2=0}^{N_c} \sum_{a_2=0}^{N_f-3}
\tu_{i_1} \tu_{i_2} \tv_{a_1} (N_f-a_1)\tv_{a_2}[N_f-a_2]_3 \pa_{\tu_{i_1}}\pa_{\tu_{i_2}}\pa_{v_{2}}\pa_{\tv_{a_2+2}},
\\
{\cal D}_{(01|12)}=&\sum_{i_1=0}^{N_c} \sum_{i_2=0}^{N_c-1} \sum_{a_1=1}^{N_f-1} \sum_{a_2=0}^{N_f-2}
\tu_{i_1} \tu_{i_2}(N_c-i_2) \tv_{a_1} (N_f-a_1)\tv_{a_2}[N_f-a_2]_3 \pa_{\tu_{i_1}}\pa_{\tu_{i_2}}\pa_{\tv_{a_1+1}}\pa_{\tv_{a_2+2}}
\nonumber\\
&+\sum_{i_1=0}^{N_c} \sum_{i_2=0}^{N_c-1} \sum_{a_2=1}^{N_f-2}
\tu_{i_1} \tu_{i_2}(N_c-i_2) \tv_{a_1} (N_f-a_1)\tv_{a_2}[N_f-a_2]_3 \pa_{\tu_{i_1}}\pa_{\tu_{i_2}}\pa_{\tv_{2}}\pa_{\tv_{a_2+1}}
\nonumber\\
&+\sum_{i_1=0}^{N_c} \sum_{i_2=0}^{N_c-1} 
\tu_{i_1} \tu_{i_2}(N_c-i_2) \tv_{a_1} (N_f-a_1)\tv_{a_2}[N_f-a_2]_3 \pa_{\tu_{i_1}}\pa_{\tu_{i_2+1}}\pa_{\tv_{2}}\pa_{\tv_{2}}.
\end{align}
Then $\tp_4={\cal O}_4\tp_0$ with (\ref{eq:diffop_sqcd_p4}) gives the $\hbar^4$-order term in the quantum SW periods. 
Different presentations of these higher order corrections can be seen in
\cite{okubothesis}.
We can continue this procedure for higher-order terms.

\section{Conclusions and Discussion}
In this paper, we studied the quantum SW curves and the quantum SW periods of the AD theories associated with ${\cal N}=2$ $SU(N_c)$ SQCD with $N_f$ hypermultiplets. 
We have found that the quantum SW periods of the AD theories are consistent with the scaling limit of the quantum periods around the superconformal point.
For type B scaling of $N_f=2n$ ($n>1$) theories, however, we need to consider the quantum correction to the potential in the quantum SW curve.
The origin of the quantum corrections to the potential seems to depend on the flavor symmetry kept manifestly.
It could be derived this term by appropriate coordinate transformations of the quantum SW curve.

We also obtained the differential operators which relate the classical periods to the quantum corrections for the AD theories up to the fourth order in $\hbar$.
It would be interesting  to explore the quantum SW curves and the quantum SW period for the AD theories which are obtained by the scaling limit of UV gauge theories with general gauge groups and hypermultiplets.
By evaluating the classical periods in terms of generalized hypergeometric functions \cite{Masuda:1996np}, one can study the quantum periods as a function of moduli parameters.
It is interesting to study  the quantum SW periods in particular locus, which is related to the conformal 
block of the Liouville CFT \cite{Bonelli:2011aa,Gaiotto:2012sf,Nishinaka:2019nuy}, the solution to the Painlev\'e equations \cite{Bonelli:2016qwg}, multi-critical points of unitary matrix models \cite{Itoyama:2019rgp}.

It is also important to study higher order corrections to the WKB periods and their integrability structure.
For  $(A_1, A_r)$-type AD theory, using the ODE/IM correspondence \cite{Dorey:2007zx}, the Borel resummed quantum SW periods are shown to satisfy the Y-system \cite{Ito:2017ypt} and the Thermodynamic Bethe ansatz (TBA) equations \cite{Ito:2018eon}.
For the $D_r$-type, see \cite{Ito:2019llq}.
Moreover, the quantum SW periods satisfy the so-called PNP  relations (see for example \cite{Codesido:2016dld}) which generalizes the scaling relations for the prepotential \cite{Matone:1995rx,Eguchi:1995jh,Sonnenschein:1995hv} and also tell us the effective central charge of the TBA system.
The associated two-dimensional conformal field theory correspond to the one predicted by the 4d/2d correspondence\cite{Beem:2013sza,Liendo:2015ofa,Cordova:2015nma,Buican:2015ina}.
The resurgence structure of the Schr\"odinger type equations are well
studied so far, while that of the SQCD type equation has not been studied in detail.

\section*{Acknowledgments}
We would like to thank Takahiro Nishinaka, Hongfei Shu and  Takahiro Uetoko for useful discussions.
The work of K.I. is supported in part by Grant-in-Aid for Scientific Research 18K03643 and 17H06463 from Japan Society for the Promotion of Science (JSPS). 

\appendix
\renewcommand{\theequation}{\Alph{section}.\arabic{equation}}
\setcounter{equation}{0}
\section{Formulas for $p_4$ in SQCD type quantum SW curve}
In this appendix,  we present some formulas for ${\cal D}_{(n_1 n_2\cdots | m_1 m_2\cdots)}$ which is necessary  the computation of $p_4$:
\begin{align}
    {\cal D}_{(0|4)}p_0=&\frac{1}{2}
    {\pa_x^4 G C \over y^3},
    \\
   {\cal D}_{(00|4)}p_0=&
   -{3\over 2}
   {C^3 \pa_x^4 G \over y^5},
   \\
  {\cal D}_{(001|3)}p_0 =&{15\over2}{C^4 \pa_x C \pa_x^3 G\over y^7}-{3\over2} {C^2\pa_x C \pa_x^3 G\over y^5},
  \\
  {\cal D}_{(01|3)}p_0
  =&-{3\over2} {C^2 \pa_x C \pa_x^3 G\over y^5},
  \\
  {\cal D}_{(1|3)}p_0=&{1\over 2}
  {\pa_x C \pa_x^3 G\over y^3},
  \\
  {\cal D}_{(011|2)}p_0=&{15\over2}{C^3 (\pa_x C)^2 \pa_x^2 G\over y^7}-{3\over2}{C (\pa_x C)^2 \pa_x^2 G\over y^5},
  \\
  {\cal D}_{(11|2)}p_0
  =&-{3\over 2}{C (\pa_x C)^2 \pa_x^2 G\over y^5},
\end{align}

\begin{align}
   {\cal D}_{(004|2)}p_0=& {15\over2}{C^4 \pa_x^4 C\pa_x^2 G\over y^7}-{3\over2} {C^2 \pa_x^4 C\pa_x^2 G\over y^5},
   \\
   {\cal D}_{(04|2)}p_0=&
   -{3\over 2}{C^2 \pa_x^4 C\pa_x^2 G\over y^5},
\\
{\cal D}_{(2|2)}p_0=&
{1\over 2}{\pa_x^2 C \pa_x^2 G\over y^3},
\\
{\cal D}_{(00|13)}p_0=&-{15\over4}{C^3 \pa_x G \pa_x^3 G\over y^7},
\\
{\cal D}_{(01|12)}p_0=&-{15\over4}{C^2 \pa_x C \pa_x G \pa_x^2 G\over y^7},
\end{align}

\begin{align}
    {\cal D}_{(0022|)}p_0=&-15{C^5 (\pa_x^2 C)^2\over y^7}+9{C^3 (\pa_x^2 C)^2 \over y^5}, \\
    {\cal D}_{(022|)}p_0=&
    3{C^3 (\pa_x^2 C)^2 \over y^5}-{C (\pa_x^2 C)^2 \over y^3},\\
    {\cal D}_{(22|)}p_0=&-{C (\pa_x^2 C)^2 \over y^3},
    \\
    {\cal D}_{(004|)}p_0=&
    3{C^4 \pa_x^4 C\over y^5}-{C^2 \pa_x^4 C \over y^3},
    \\
    {\cal D}_{(04|)}p_0=&{C^2 \pa_x^4 C \over y^3},
    \\
{\cal D}_{(4|)}p_0=&{\pa_x^4 C\over y}.
\end{align}

\section{Quantum SW periods for AD theories associated with $SU(2)$ and $SU(3)$ SQCDs}

In this appendix,
we write down  quantum correction to the SW periods for the AD theories
obtained by  the scaling limit of $SU(2)$ and $SU(3)$ SQCDs.
\subsection{$SU(2)$}
For the AD theories associated with $SU(2)$ \cite{Argyres:1995xn}, the scaling limit of the quantum corrections has been studied in \cite{Ito:2018eon}.
For $N_f=1$, from (\ref{eq:nf01p2}) and (\ref{eq:nf01p4}) we obtain
\begin{align}
    \tilde{p}_2=&
    \frac{1}{4}\partial_{\tilde{u}_2}\partial_{\tilde{u}_3}\tilde{p}_0,
    \\
    \tilde{p}_4=&
    \frac{7}{160}\partial_{\tilde{u}_2}^2\partial_{\tilde{u}_3}^2\tilde{p}_0.
\end{align}
For $N_f=2$, from (\ref{eq:nf2p2}) and (\ref{eq:nf2p4}), we have
\begin{align}
    \tilde{p}_2=&
    {1\over12} \sum_{i=0}^{2} \tilde{u}_i (3-i)^2 \partial_{\tilde{u}_{i+1}}\partial_{\tilde{u}_{2}}\tilde{p}_0,
    \\
    \tilde{p}_4=& {7\over 1440} \sum_{i=0}^{2} \sum_{j=0}^{2}
\tilde{u}_i \tilde{u}_j (3-i)^2(3-j)^2 \partial_{\tilde{u}_{i+1}}\partial_{\tilde{u}_{j+1}}\partial_{\tilde{u}_{2}}^2 \tp_0
-{1\over288} \sum_{i=0}^{2} \tilde{u}_i (3-i)^4 \partial_{\tilde{u}_{i+1}}\partial_{\tilde{u}_{2}}\partial_{\tilde{u}_3}\tp_0.
\end{align}

For $N_f=3$, we apply (\ref{eq:p2_sqcd}) and (\ref{eq:p4_sqcd}).
We find 
\begin{align}
\tp_2=&{1\over24} 
\left(
\sum_{i=0}^{2}  2\tu_i \tu_0 
\partial_{\tu_i}  \partial_{\tu_{2}}
+ 2\tu_0    \partial_{\tu_2}
\right)\tp_0
-{1\over24} \sum_{i=0}^1  \tu_i (2-i) \left(
3 \partial_{\tu_i} \partial_{\tv_2} 
+ \tv_2 \partial_{\tu_{i+1}}\partial_{\tv_{a+1}} 
\right)\tp_0
\nonumber\\
&
+{1\over24}\sum_{i=0}^{2} \sum_{a=0}^{1}  \tu_i \tv_a (3-a)(2-a)
 \partial_{\tu_i} \partial_{\tv_{a+2}}
\tp_0.
\end{align}
For $\tilde{p}_4$, we have
$\pa_{\tx}^3 \tC=0$ and  $\pa_{\tx}^4 \tG=0$.
Then we can set
\begin{align}
{\cal D}_{(0|4)}\tp_0=&{\cal D}_{(00|4)}\tp_0={\cal D}_{(004|2)}\tp_0=
{\cal D}_{(04|2)}\tp_0={\cal D}_{(004|)}\tp_0={\cal D}_{(04|)}\tp_0={\cal D}_{(4|)}\tp_0=0,
\end{align}
in (\ref{eq:diffop_sqcd_p4}). 
We then find
\begin{align}
\tilde{p}_4=&\left\{
{11\over 34560} {\cal D}_{(001|3)}
+{19\over 8640}  {\cal D}_{(01|3)}+{1\over 360} {\cal D}_{(1|3)}
\right.
+{7\over 5760}  {\cal D}_{(011|2)}
+{7\over 1440}  {\cal D}_{(11|2)}
\nonumber\\
&
+{19\over 1920}  {\cal D}_{(2|2)}
+{11\over 34560} 
{\cal D}_{(00|13)}
-{7\over 5760} {\cal D}_{(01|12)}
+{7\over 5760} {\cal D}_{(0022|)}
\nonumber\\
&+\left.
{37\over 5760 }  {\cal D}_{(022|)}
+{1\over 180} {\cal D}_{(22|)}
\right\}
\tilde{p}_0.
\end{align}

\subsection{$SU(3)$ case}
For $N_f=0$, from (\ref{eq:nf01p2}) and (\ref{eq:nf01p4}), we get
\begin{align}
    \tilde{p}_2=&
    \frac{1}{4}\partial_{\tilde{u}_2}\partial_{\tilde{u}_3}\tilde{p}_0,
    \\
    \tilde{p}_4=&
    \frac{7}{160}\partial_{\tilde{u}_2}^2\partial_{\tilde{u}_3}^2\tilde{p}_0.
\end{align}
For $N_f=1$ case, we have
\begin{align}
    \tilde{p}_2=&
    \left(
    \frac{1}{2}\partial_{\tilde{u}_2}\partial_{\tilde{u}_4}
    -\frac{1}{6}\tilde{u}_2\partial_{\tilde{u}_4}^2
    \right)\tilde{p}_0,
    \\
    \tilde{p}_4=&
    \left(
    \frac{7}{40}\partial_{\tilde{u}_2}^2\partial_{\tilde{u}_4}^2
    -\frac{1}{48}\partial_{\tilde{u}_4}^3
    \right)\tilde{p}_0.
\end{align}
For $N_f=2$,  (\ref{eq:nf2p2}) and (\ref{eq:nf2p4}) lead to
\begin{align}
\tilde{p}_2=&{1\over12} \sum_{i=0}^{3} \tilde{u}_i (4-i)^2 \partial_{\tilde{u}_{i+1}}\partial_{\tilde{u}_{3}}\tilde{p}_0,
\end{align}
and
\begin{align}
\tp_4=& {7\over 1440} \sum_{i=0}^{3} \sum_{j=0}^{3}
\tilde{u}_i \tilde{u}_j (4-i)^2(4-j)^2 \partial_{\tilde{u}_{i+1}}\partial_{\tilde{u}_{j+1}}\partial_{\tilde{u}_{3}}^2 \tp_0
-{1\over288} \sum_{i=0}^{3} \tilde{u}_i (4-i)^4 \partial_{\tilde{u}_{i+1}}\partial_{\tilde{u}_{3}}\partial_{\tilde{u}_{4}}\tp_0.
\end{align}
For $N_f=3$, from (\ref{eq:p2_sqcd}), we obtain
\begin{align}
\tp_2=&{1\over24} 
\left(
\sum_{i=0}^{3} \sum_{j=0}^{1} \tu_i \tu_j (3-j)(2-j)  \partial_{\tu_i}  \partial_{\tu_{j+2}}
+\sum_{j=0}^{1} \tu_j (3-j)(2-j)   \partial_{\tu_{j+2}}
\right)\tp_0
\nonumber\\
&
-{1\over24} \sum_{i=0}^{2}  \tu_i (3-i) \left(
3 \partial_{\tu_i} \partial_{\tv_2} 
+ \tv_2  \partial_{\tu_{i+1}}\partial_{\tv_3 }
\right)\tp_0
+{1\over24}\sum_{i=0}^3 \sum_{a=0}^{1}  \tu_i \tv_a (3-a)(2-a)
 \partial_{\tu_i} \partial_{\tv_{a+2}}
\tp_0.
\end{align}
For $\tp_4$,  
$\pa_{\tx}^3 \tC=0$ and $\pa_{\tx}^4 \tG=0$ hold.
Then we can set
\begin{align}
{\cal D}_{(0|4)}\tp_0=&{\cal D}_{(00|4)}\tp_0={\cal D}_{(004|2)}\tp_0=
{\cal D}_{(04|2)}\tp_0={\cal D}_{(004|)}\tp_0={\cal D}_{(04|)}\tp_0={\cal D}_{(4|)}\tp_0=0.
\end{align}
Therefore, we get
\begin{align}
\tilde{p}_4=&
\left\{
+{11\over 34560} {\cal D}_{(001|3)}
+{19\over 8640}  {\cal D}_{(01|3)}+{1\over 360} {\cal D}_{(1|3)}\right.
+{7\over 5760}  {\cal D}_{(011|2)}
+{7\over 1440}  {\cal D}_{(11|2)}
\nonumber\\
&
+{19\over 1920}  {\cal D}_{(2|2)}
+{11\over 34560} 
{\cal D}_{(00|13)}
-{7\over 5760} {\cal D}_{(01|12)}
+{7\over 5760} {\cal D}_{(0022|)}
\nonumber\\
&+\left.{37\over 5760 }  {\cal D}_{(022|)}
+{1\over 180} {\cal D}_{(22|)}\right\}\tilde{p}_0.
\end{align}

For $N_f=4$ case, there are two types of the scaling limits.
For the case of type A scaling,
$\tp_2$ is given by
\begin{align}
    \tilde{p}_2=&
    \frac{1}{24}\left(
    \sum_{i=0}^4\sum_{j=0}^2\tu_i\tu_j(4-j)(3-j)\pa_{\tu_i}\pa_{\tu_{j+2}}
    +\sum_{j=0}^2\tu_j(4-j)(3-j)\pa_{\tu_{j+2}}\right)\tilde{p}_0
    \notag \\
    &-\frac{1}{24}\sum_{i=0}^3\tu_i(4-i)
    \left(
    4\pa_{\tu_i}\pa_{\tv_2}+\sum_{a=2}^3\tv_a(4-a)\pa_{\tu_{i+1}}\pa_{\tv_{a+1}}\right)\tilde{p}_0
    \notag \\
    &+\frac{1}{24}\sum_{i=0}^4\sum_{a=0}^2
    \tu_i\tv_a(4-a)(3-a)\pa_{\tu_j}\pa_{\tv_{a+2}}\tilde{p}_0.
\end{align}
For $\tp_4$, we have
$
\pa_{\tx}^4 \tC=0
$.
Then  we can set 
\begin{align}
{\cal D}_{(004|2)}\tp_0=
{\cal D}_{(04|2)}\tp_0= {\cal D}_{(004|)}\tp_0={\cal D}_{(04|)}\tp_0= {\cal D}_{(4|)}\tp_0=0.
\end{align}
We obtain
\begin{align}
\tp_4
=&\left\{{19\over 5760}{\cal D}_{(0|4)}+{41 \over 34560}{\cal D}_{(00|4)}
+{11\over 34560} {\cal D}_{(001|3)}
+{19\over 8640}  {\cal D}_{(01|3)}+{1\over 360} {\cal D}_{(1|3)}
\right.
\nonumber\\
&
+{7\over 5760}  {\cal D}_{(011|2)}
+{7\over 1440}  {\cal D}_{(11|2)}
+{19\over 1920}  {\cal D}_{(2|2)}
+{11\over 34560} 
{\cal D}_{(00|13)}
-{7\over 5760} {\cal D}_{(01|12)}
\nonumber\\
&+{7\over 5760} {\cal D}_{(0022|)}
+\left.{37\over 5760 }  {\cal D}_{(022|)}
+{1\over 180} {\cal D}_{(22|)}\right\}\tp_0.
\end{align}
For  type B scaling, from (\ref{eq:p2nfeventypeb}) and (\ref{eq:p4nfeventypeb}),
we get
\begin{align}
    \tilde{p}_2=&
    \frac{1}{12}\sum_{i=0}^3\tu_i(i-5)(i-6)\pa_{\tu_{i+1}}\pa_{\tu_2}\tilde{p}_0
    +\frac{1}{12}\tu_36\pa_{\tu_3}\pa_{\tu_3}\tilde{p}_0
    +\frac{1}{2}\pa_{\tu_3}\tilde{p}_0,\\
\tp_4=&{7\over 1440}\Bigl\{
\sum_{i=0}^{2}\sum_{j=0}^{2}
\tu_i \tu_j (i-5)(i-6)(j-5)(j-6) \partial_{\tu_{i+1}}\partial_{\tu_{j+1}}\partial_{\tu_2}^2 p_0
\nonumber\\
&+12 \sum_{i=0}^{2}\tu_i \tu_2 (i-5)(i-6)
 \partial_{\tu_{i+1}} \partial_{\tu_3} \partial_{\tu_2} \partial_{\tu_3} \tp_0
+36\tu_3^2 
\partial_{\tu_{3}}^4 
\tp_0
\Bigr\}
\nonumber\\
&
+{1\over 288} \Bigl\{ \sum_{i=0}^2 \tu_i (i-5)(i-6)(i-7)(i-8)
\partial_{\tu_{i+1}}\partial_{\tu_{3}}\partial_{\tu_{2}}\tp_0
+120\tu_3 
\partial_{\tu_3}\partial_{\tu_{3}}^2 \tp_0
\Bigr\}
\nonumber\\
&
-{1\over64} \Bigl\{ \sum_{i=0}^{2} \tu_i (i-5)(i-6)
\partial_{\tu_{i+1}}\partial_{\tu_{3}}\partial_{\tu_{2}}\tp_0
+6\tu_{3} 
\partial_{\tu_{3}}\partial_{\tu_{3}}^2 \tp_0
\Bigr\}
+\frac{3}{8}\partial_{\tu_{3}}^2\tp_0.
\end{align}
Finally, for
$N_f=5$, $\tp_2$ is given by
\begin{align}
\tp_2=&{1\over24} 
\left(
\sum_{i=0}^{3} \sum_{j=0}^{1} \tu_i \tu_j (3-j)(2-j)  \partial_{\tu_i}  \partial_{\tu_{j+2}}
+\sum_{j=0}^{1} \tu_j (3-j)(2-j)   \partial_{\tu_{j+2}}
\right)\tp_0
\nonumber\\
&
-{1\over24} \sum_{i=0}^{2}  \tu_i (3-i) \left(
N_f \partial_{\tu_i} \partial_{\tv_2} +\sum_{a=2}^{4} \tv_a (5-a) \partial_{\tu_{i+1}}\partial_{\tv_{a+1}} 
\right)\tp_0
\nonumber\\
&
+{1\over24}\sum_{i=0}^{3} \sum_{a=0}^{3}  \tu_i \tv_a (5-a)(4-a)
 \partial_{\tu_i} \partial_{\tv_{a+2}}
\tp_0.
\end{align}
For $\tp_4$, we have
$
\pa_{\tx}^4 \tC=0
$.
Then, by setting 
\begin{align}
{\cal D}_{(004|2)}\tp_0=
{\cal D}_{(04|2)}\tp_0={\cal D}_{(004|)}\tp_0={\cal D}_{(04|)}\tp_0={\cal D}_{(4|)}\tp_0=0.
\end{align}
we have
\begin{align}
\tp_4
=&\left\{{19\over 5760}{\cal D}_{(0|4)}+{41 \over 34560}{\cal D}_{(00|4)}
+{11\over 34560} {\cal D}_{(001|3)}
+{19\over 8640}  {\cal D}_{(01|3)}+{1\over 360} {\cal D}_{(1|3)}\right.
\nonumber\\
&+{7\over 5760}  {\cal D}_{(011|2)}
+{7\over 1440}  {\cal D}_{(11|2)}
+{19\over 1920}  {\cal D}_{(2|2)}
+{11\over 34560} 
{\cal D}_{(00|13)}
\nonumber\\
&
-{7\over 5760} {\cal D}_{(01|12)}
+{7\over 5760} {\cal D}_{(0022|)}
\left.+{37\over 5760 }  {\cal D}_{(022|)}
+{1\over 180} {\cal D}_{(22|)}\right\}\tp_0.
\end{align}


\begin{thebibliography}{10}

\bibitem{Seiberg:1994rs}
N.~Seiberg and E.~Witten,  ``{Electric - magnetic duality, monopole
  condensation, and confinement in N=2 supersymmetric Yang-Mills
  theory},''\href{https://doi.org/10.1016/0550-3213(94)90124-4,
  10.1016/0550-3213(94)00449-8}{Nucl. Phys. {\bfseries B426} (1994) 19}
  [\href{https://arxiv.org/abs/hep-th/9407087}{{
  hep-th/9407087}}].

\bibitem{Seiberg:1994aj}
N.~Seiberg and E.~Witten,  ``{Monopoles, duality and chiral symmetry breaking
  in N=2 supersymmetric
  QCD},''\href{https://doi.org/10.1016/0550-3213(94)90214-3}{Nucl. Phys.
  {\bfseries B431} (1994) 484}
  [\href{https://arxiv.org/abs/hep-th/9408099}{{
  hep-th/9408099}}].

\bibitem{Argyres:1995jj}
P.~C. Argyres and M.~R. Douglas,  ``{New phenomena in SU(3) supersymmetric
  gauge theory},''\href{https://doi.org/10.1016/0550-3213(95)00281-V}{Nucl.
  Phys. {\bfseries B448} (1995) 93}
  [\href{https://arxiv.org/abs/hep-th/9505062}{{
  hep-th/9505062}}].

\bibitem{Argyres:1995xn}
P.~C. Argyres, M.~R. Plesser, N.~Seiberg and E.~Witten,  ``{New N=2
  superconformal field theories in
  four-dimensions},''\href{https://doi.org/10.1016/0550-3213(95)00671-0}{Nucl.
  Phys. {\bfseries B461} (1996) 71}
  [\href{https://arxiv.org/abs/hep-th/9511154}{{
  hep-th/9511154}}].

\bibitem{Eguchi:1996vu}
T.~Eguchi, K.~Hori, K.~Ito and S.-K. Yang,  ``{Study of N=2 superconformal
  field theories in
  four-dimensions},''\href{https://doi.org/10.1016/0550-3213(96)00188-5}{Nucl.
  Phys. {\bfseries B471} (1996) 430}
  [\href{https://arxiv.org/abs/hep-th/9603002}{{
  hep-th/9603002}}].

\bibitem{Gaiotto:2009we}
D.~Gaiotto,  ``{N=2
  dualities},''\href{https://doi.org/10.1007/JHEP08(2012)034}{JHEP {\bfseries
  08} (2012) 034} [\href{https://arxiv.org/abs/0904.2715}{{
  0904.2715}}].

\bibitem{Gaiotto:2009hg}
D.~Gaiotto, G.~W. Moore and A.~Neitzke,  ``{Wall-crossing, Hitchin Systems, and
  the WKB Approximation},'' \href{https://arxiv.org/abs/0907.3987}{{0907.3987}}.

\bibitem{Xie:2012hs}
D.~Xie,  ``{General Argyres-Douglas
  Theory},''\href{https://doi.org/10.1007/JHEP01(2013)100}{JHEP {\bfseries 01}
  (2013) 100} [\href{https://arxiv.org/abs/1204.2270}{{
  1204.2270}}].

\bibitem{Cecotti:2010fi}
S.~Cecotti, A.~Neitzke and C.~Vafa,  ``{R-Twisting and 4d/2d
  Correspondences},'' \href{https://arxiv.org/abs/1006.3435}{{1006.3435}}.

\bibitem{Wang:2015mra}
Y.~Wang and D.~Xie,  ``{Classification of Argyres-Douglas theories from M5
  branes},''\href{https://doi.org/10.1103/PhysRevD.94.065012}{Phys. Rev.
  {\bfseries D94} (2016) 065012} [\href{https://arxiv.org/abs/1509.00847}{{1509.00847}}].

\bibitem{Wang:2016yha}
Y.~Wang, D.~Xie, S.~S.~T. Yau and S.-T. Yau,  ``{$4d$ $\mathcal{N} = 2$ SCFT
  from complete intersection
  singularity},''\href{https://doi.org/10.4310/ATMP.2017.v21.n3.a6}{Adv. Theor.
  Math. Phys. {\bfseries 21} (2017) 801}
  [\href{https://arxiv.org/abs/1606.06306}{{\if0 \ttfamily \fi 1606.06306}}].

\bibitem{Argyres:2015ffa}
P.~Argyres, M.~Lotito, Y.~Lu and M.~Martone,  ``{Geometric constraints on the
  space of $ \mathcal{N} $ = 2 SCFTs. Part I: physical constraints on relevant
  deformations},''\href{https://doi.org/10.1007/JHEP02(2018)001}{JHEP
  {\bfseries 02} (2018) 001} [\href{https://arxiv.org/abs/1505.04814}{{1505.04814}}].

\bibitem{Nekrasov:2009rc}
N.~A. Nekrasov and S.~L. Shatashvili,  ``{Quantization of Integrable Systems
  and Four Dimensional Gauge Theories},''
  \href{https://arxiv.org/abs/0908.4052}{{\if0 \ttfamily \fi 0908.4052}}.

\bibitem{Moore:1997dj}
G.~W. Moore, N.~Nekrasov and S.~Shatashvili,  ``{Integrating over Higgs
  branches},''\href{https://doi.org/10.1007/PL00005525}{Commun. Math. Phys.
  {\bfseries 209} (2000) 97} [\href{https://arxiv.org/abs/hep-th/9712241}{{hep-th/9712241}}].

\bibitem{Mironov:2009uv}
A.~Mironov and A.~Morozov,  ``{Nekrasov Functions and Exact Bohr-Zommerfeld
  Integrals},''\href{https://doi.org/10.1007/JHEP04(2010)040}{JHEP {\bfseries
  04} (2010) 040} [\href{https://arxiv.org/abs/0910.5670}{{
  0910.5670}}].

\bibitem{Basar:2015xna}
G.~Ba\c{s}ar and G.~V. Dunne,  ``{Resurgence and the Nekrasov-Shatashvili
  limit: connecting weak and strong coupling in the Mathieu and Lame
  systems},''\href{https://doi.org/10.1007/JHEP02(2015)160}{JHEP {\bfseries 02}
  (2015) 160} [\href{https://arxiv.org/abs/1501.05671}{{
  1501.05671}}].

\bibitem{Kashani-Poor:2015pca}
A.-K. Kashani-Poor and J.~Troost,  ``{Pure $ \mathcal{N}=2 $ super Yang-Mills
  and exact WKB},''\href{https://doi.org/10.1007/JHEP08(2015)160}{JHEP
  {\bfseries 08} (2015) 160} [\href{https://arxiv.org/abs/1504.08324}{{1504.08324}}].

\bibitem{Ashok:2016yxz}
S.~K. Ashok, D.~P. Jatkar, R.~R. John, M.~Raman and J.~Troost,  ``{Exact WKB
  analysis of $ \mathcal{N} $ = 2 gauge
  theories},''\href{https://doi.org/10.1007/JHEP07(2016)115}{JHEP {\bfseries
  07} (2016) 115} [\href{https://arxiv.org/abs/1604.05520}{{
  1604.05520}}].

\bibitem{Mironov:2009dv}
A.~Mironov and A.~Morozov,  ``{Nekrasov Functions from Exact BS Periods: The
  Case of SU(N)},''\href{https://doi.org/10.1088/1751-8113/43/19/195401}{J.
  Phys. {\bfseries A43} (2010) 195401}
  [\href{https://arxiv.org/abs/0911.2396}{{\if0 \ttfamily \fi 0911.2396}}].

\bibitem{Zenkevich:2011zx}
Y.~Zenkevich,  ``{Nekrasov prepotential with fundamental matter from the
  quantum spin
  chain},''\href{https://doi.org/10.1016/j.physletb.2011.06.030}{Phys. Lett.
  {\bfseries B701} (2011) 630} [\href{https://arxiv.org/abs/1103.4843}{{1103.4843}}].

\bibitem{Popolitov:2013ria}
A.~V. Popolitov,  ``{Relation between Nekrasov functions and Bohr-Sommerfeld
  periods in the pure $SU(N)$
  case},''\href{https://doi.org/10.1007/s11232-014-0139-0}{Theor. Math. Phys.
  {\bfseries 178} (2014) 239}.

\bibitem{He:2010xa}
W.~He and Y.-G. Miao,  ``{Magnetic expansion of Nekrasov theory: the SU(2) pure
  gauge theory},''\href{https://doi.org/10.1103/PhysRevD.82.025020}{Phys. Rev.
  {\bfseries D82} (2010) 025020} [\href{https://arxiv.org/abs/1006.1214}{{1006.1214}}].

\bibitem{Ito:2017iba}
K.~Ito, S.~Kanno and T.~Okubo,  ``{Quantum periods and prepotential in $
  \mathcal{N}=2 $ SU(2)
  SQCD},''\href{https://doi.org/10.1007/JHEP08(2017)065}{JHEP {\bfseries 08}
  (2017) 065} [\href{https://arxiv.org/abs/1705.09120}{{
  1705.09120}}].

\bibitem{Beccaria:2016wop}
M.~Beccaria,  ``{On the large $\Omega$-deformations in the Nekrasov-Shatashvili
  limit of $\mathcal N=2^{*}$
  SYM},''\href{https://doi.org/10.1007/JHEP07(2016)055}{JHEP {\bfseries 07}
  (2016) 055} [\href{https://arxiv.org/abs/1605.00077}{{
  1605.00077}}].

\bibitem{Huang:2012kn}
M.-x. Huang,  ``{On Gauge Theory and Topological String in Nekrasov-Shatashvili
  Limit},''\href{https://doi.org/10.1007/JHEP06(2012)152}{JHEP {\bfseries 06}
  (2012) 152} [\href{https://arxiv.org/abs/1205.3652}{{
  1205.3652}}].

\bibitem{Krefl:2014nfa}
D.~Krefl,  ``{Non-Perturbative Quantum Geometry
  II},''\href{https://doi.org/10.1007/JHEP12(2014)118}{JHEP {\bfseries 12}
  (2014) 118} [\href{https://arxiv.org/abs/1410.7116}{{
  1410.7116}}].

\bibitem{Maruyoshi:2010iu}
K.~Maruyoshi and M.~Taki,  ``{Deformed Prepotential, Quantum Integrable System
  and Liouville Field
  Theory},''\href{https://doi.org/10.1016/j.nuclphysb.2010.08.008}{Nucl. Phys.
  {\bfseries B841} (2010) 388} [\href{https://arxiv.org/abs/1006.4505}{{1006.4505}}].

\bibitem{Kanno:2013vi}
H.~Kanno, K.~Maruyoshi, S.~Shiba and M.~Taki,  ``{$W_3$ irregular states and
  isolated N=2 superconformal field
  theories},''\href{https://doi.org/10.1007/JHEP03(2013)147}{JHEP {\bfseries
  03} (2013) 147} [\href{https://arxiv.org/abs/1301.0721}{{
  1301.0721}}].

\bibitem{Piatek:2014lma}
M.~Piatek and A.~R. Pietrykowski,  ``{Classical irregular block, $ \mathcal{N}
  $ = 2 pure gauge theory and Mathieu
  equation},''\href{https://doi.org/10.1007/JHEP12(2014)032}{JHEP {\bfseries
  12} (2014) 032} [\href{https://arxiv.org/abs/1407.0305}{{
  1407.0305}}].

\bibitem{Poghossian:2016rzb}
R.~Poghossian,  ``{Deformed SW curve and the null vector decoupling equation in
  Toda field theory},''\href{https://doi.org/10.1007/JHEP04(2016)070}{JHEP
  {\bfseries 04} (2016) 070} [\href{https://arxiv.org/abs/1601.05096}{{1601.05096}}].

\bibitem{Ito:2018hwp}
K.~Ito and T.~Okubo,  ``{Quantum periods for $\mathcal{N}=2$ $SU(2)$ SQCD
  around the superconformal
  point},''\href{https://doi.org/10.1016/j.nuclphysb.2018.07.007}{Nucl. Phys.
  {\bfseries B934} (2018) 356} [\href{https://arxiv.org/abs/1804.04815}{{1804.04815}}].

\bibitem{Ito:2019twh}
K.~Ito, S.~Koizumi and T.~Okubo,  ``{Quantum Seiberg-Witten curve and
  Universality in Argyres-Douglas
  theories},''\href{https://doi.org/10.1016/j.physletb.2019.03.024}{Phys. Lett.
  {\bfseries B792} (2019) 29} [\href{https://arxiv.org/abs/1903.00168}{{1903.00168}}].

\bibitem{Gaiotto:2014bza}
D.~Gaiotto,  ``{Opers and TBA},'' \href{https://arxiv.org/abs/1403.6137}{{1403.6137}}.

\bibitem{Ito:2018eon}
K.~Ito, M.~Marino and H.~Shu,  ``{TBA equations and resurgent Quantum
  Mechanics},''\href{https://doi.org/10.1007/JHEP01(2019)228}{JHEP {\bfseries
  01} (2019) 228} [\href{https://arxiv.org/abs/1811.04812}{{
  1811.04812}}].

\bibitem{Ito:2017ypt}
K.~Ito and H.~Shu,  ``{ODE/IM correspondence and the Argyres-Douglas
  theory},''\href{https://doi.org/10.1007/JHEP08(2017)071}{JHEP {\bfseries 08}
  (2017) 071} [\href{https://arxiv.org/abs/1707.03596}{{
  1707.03596}}].

\bibitem{Grassi:2019coc}
A.~Grassi, J.~Gu and M.~Marino,  ``{Non-perturbative approaches to the quantum
  Seiberg-Witten curve},'' \href{https://arxiv.org/abs/1908.07065}{{1908.07065}}.

\bibitem{Fioravanti:2019vxi}
D.~Fioravanti and D.~Gregori,  ``{Integrability and cycles of deformed ${\cal
  N}=2$ gauge theory},'' \href{https://arxiv.org/abs/1908.08030}{{1908.08030}}.

\bibitem{Ito:2019llq}
K.~Ito and H.~Shu,  ``{TBA equations for the Schr\"odinger equation with a
  regular singularity},'' \href{https://arxiv.org/abs/1910.09406}{{1910.09406}}.

\bibitem{Gaiotto:2010jf}
D.~Gaiotto, N.~Seiberg and Y.~Tachikawa,  ``{Comments on scaling limits of 4d
  N=2 theories},''\href{https://doi.org/10.1007/JHEP01(2011)078}{JHEP
  {\bfseries 01} (2011) 078} [\href{https://arxiv.org/abs/1011.4568}{{1011.4568}}].

\bibitem{Hanany:1995na}
A.~Hanany and Y.~Oz,  ``{On the quantum moduli space of vacua of N=2
  supersymmetric SU(N(c)) gauge
  theories},''\href{https://doi.org/10.1016/0550-3213(95)00376-4}{Nucl. Phys.
  {\bfseries B452} (1995) 283}
  [\href{https://arxiv.org/abs/hep-th/9505075}{{
  hep-th/9505075}}].

\bibitem{Cardy:1988cwa}
J.~L. Cardy,  ``{Is There a c Theorem in
  Four-Dimensions?},''\href{https://doi.org/10.1016/0370-2693(88)90054-8}{Phys.
  Lett. {\bfseries B215} (1988) 749}.

\bibitem{Brandhuber:1995zp}
A.~Brandhuber and K.~Landsteiner,  ``{On the monodromies of N=2 supersymmetric
  Yang-Mills theory with gauge group
  SO(2n)},''\href{https://doi.org/10.1016/0370-2693(95)00986-U}{Phys. Lett.
  {\bfseries B358} (1995) 73}
  [\href{https://arxiv.org/abs/hep-th/9507008}{{
  hep-th/9507008}}].

\bibitem{okubothesis}
T.~Okubo,  Quantum periods for N = 2 Supersymmetric QCD at strong
  coupling, Ph.D. thesis, Tokyo Institute of Technology, 2018.

\bibitem{Masuda:1996np}
T.~Masuda and H.~Suzuki,  ``{On explicit evaluations around the conformal point
  in N=2 supersymmetric Yang-Mills
  theories},''\href{https://doi.org/10.1016/S0550-3213(97)00199-5}{Nucl. Phys.
  {\bfseries B495} (1997) 149}
  [\href{https://arxiv.org/abs/hep-th/9612240}{{
  hep-th/9612240}}].

\bibitem{Bonelli:2011aa}
G.~Bonelli, K.~Maruyoshi and A.~Tanzini,  ``{Wild Quiver Gauge
  Theories},''\href{https://doi.org/10.1007/JHEP02(2012)031}{JHEP {\bfseries
  1202} (2012) 031} [\href{https://arxiv.org/abs/1112.1691}{{
  1112.1691}}].

\bibitem{Gaiotto:2012sf}
D.~Gaiotto and J.~Teschner,  ``{Irregular singularities in Liouville theory and
  Argyres-Douglas type gauge theories,
  I},''\href{https://doi.org/10.1007/JHEP12(2012)050}{JHEP {\bfseries 12}
  (2012) 050} [\href{https://arxiv.org/abs/1203.1052}{{
  1203.1052}}].

\bibitem{Nishinaka:2019nuy}
T.~Nishinaka and T.~Uetoko,  ``{Argyres-Douglas theories and Liouville
  Irregular States},''\href{https://doi.org/10.1007/JHEP09(2019)104}{JHEP
  {\bfseries 09} (2019) 104} [\href{https://arxiv.org/abs/1905.03795}{{1905.03795}}].

\bibitem{Bonelli:2016qwg}
G.~Bonelli, O.~Lisovyy, K.~Maruyoshi, A.~Sciarappa and A.~Tanzini,  ``{On
  Painlev\'e/gauge theory correspondence},''
  \href{https://arxiv.org/abs/1612.06235}{{\if0 \ttfamily \fi 1612.06235}}.

\bibitem{Itoyama:2019rgp}
H.~Itoyama, T.~Oota and K.~Yano,  ``{Multicritical points of unitary matrix
  model with logarithmic potential identified with Argyres-Douglas points},''
  \href{https://arxiv.org/abs/1909.10770}{{\if0 \ttfamily \fi 1909.10770}}.

\bibitem{Dorey:2007zx}
P.~Dorey, C.~Dunning and R.~Tateo,  ``{The ODE/IM
  Correspondence},''\href{https://doi.org/10.1088/1751-8113/40/32/R01}{J. Phys.
  {\bfseries A40} (2007) R205}
  [\href{https://arxiv.org/abs/hep-th/0703066}{{
  hep-th/0703066}}].

\bibitem{Codesido:2016dld}
S.~Codesido and M.~Marino,  ``{Holomorphic Anomaly and Quantum
  Mechanics},''\href{https://doi.org/10.1088/1751-8121/aa9e77}{J. Phys.
  {\bfseries A51} (2018) 055402} [\href{https://arxiv.org/abs/1612.07687}{{1612.07687}}].

\bibitem{Matone:1995rx}
M.~Matone,  ``{Instantons and recursion relations in N=2 SUSY gauge
  theory},''\href{https://doi.org/10.1016/0370-2693(95)00920-G}{Phys. Lett.
  {\bfseries B357} (1995) 342}
  [\href{https://arxiv.org/abs/hep-th/9506102}{{
  hep-th/9506102}}].

\bibitem{Eguchi:1995jh}
T.~Eguchi and S.-K. Yang,  ``{Prepotentials of N=2 supersymmetric gauge
  theories and soliton
  equations},''\href{https://doi.org/10.1142/S0217732396000151}{Mod. Phys.
  Lett. {\bfseries A11} (1996) 131}
  [\href{https://arxiv.org/abs/hep-th/9510183}{{
  hep-th/9510183}}].

\bibitem{Sonnenschein:1995hv}
J.~Sonnenschein, S.~Theisen and S.~Yankielowicz,  ``{On the relation between
  the holomorphic prepotential and the quantum moduli in SUSY gauge
  theories},''\href{https://doi.org/10.1016/0370-2693(95)01399-7}{Phys. Lett.
  {\bfseries B367} (1996) 145}
  [\href{https://arxiv.org/abs/hep-th/9510129}{{
  hep-th/9510129}}].

\bibitem{Beem:2013sza}
C.~Beem, M.~Lemos, P.~Liendo, W.~Peelaers, L.~Rastelli and B.~C. van Rees,
  ``{Infinite Chiral Symmetry in Four
  Dimensions},''\href{https://doi.org/10.1007/s00220-014-2272-x}{Commun. Math.
  Phys. {\bfseries 336} (2015) 1359}
  [\href{https://arxiv.org/abs/1312.5344}{{\if0 \ttfamily \fi 1312.5344}}].

\bibitem{Liendo:2015ofa}
P.~Liendo, I.~Ramirez and J.~Seo,  ``{Stress-tensor OPE in $ \mathcal{N}=2 $
  superconformal
  theories},''\href{https://doi.org/10.1007/JHEP02(2016)019}{JHEP {\bfseries
  02} (2016) 019} [\href{https://arxiv.org/abs/1509.00033}{{
  1509.00033}}].

\bibitem{Cordova:2015nma}
C.~Cordova and S.-H. Shao,  ``{Schur Indices, BPS Particles, and
  Argyres-Douglas
  Theories},''\href{https://doi.org/10.1007/JHEP01(2016)040}{JHEP {\bfseries
  01} (2016) 040} [\href{https://arxiv.org/abs/1506.00265}{{
  1506.00265}}].

\bibitem{Buican:2015ina}
M.~Buican and T.~Nishinaka,  ``{On the superconformal index of Argyres-Douglas
  theories},''\href{https://doi.org/10.1088/1751-8113/49/1/015401}{J. Phys.
  {\bfseries A49} (2016) 015401} [\href{https://arxiv.org/abs/1505.05884}{{1505.05884}}].

\end{thebibliography}


\providecommand{\href}[2]{#2}\begingroup\raggedright\endgroup

\end{document}